\documentclass[fleqn,usenatbib]{rasti}
\usepackage{newtxtext,newtxmath}
\usepackage[T1]{fontenc}

\DeclareRobustCommand{\VAN}[3]{#2}
\let\VANthebibliography\thebibliography
\def\thebibliography{\DeclareRobustCommand{\VAN}[3]{##3}\VANthebibliography}

\usepackage{graphicx}	
\usepackage{amsmath}	
\usepackage{xspace} 
\usepackage{lineno}
\usepackage{comment}

\newcommand{\be}{\begin{equation}}
\newcommand{\ee}{\end{equation}}

\newcommand{\bi}{\begin{itemize}}
\newcommand{\ei}{\end{itemize}}
\newcommand{\ben}{\begin{enumerate}}
\newcommand{\een}{\end{enumerate}}
\newcommand{\onesize}{0.95}
\newcommand{\twosize}{0.7}
\newcommand{\threesize}{0.65}
\newcommand{\cmsize}{0.85}

\newcommand{\zenodo}{\url{https://zenodo.org/doi/10.5281/zenodo.8140548}}

\newcommand{\Fermi}{{\textit{Fermi}}}

\title[Covariate shift effect on classification]{Effect of covariate shift on multi-class classification of Fermi-LAT sources}

\author[D. V. Malyshev]{
Dmitry V. Malyshev$^{1}$\thanks{E-mail: dvmalyshev@gmail.com}
\\
$^{1}$Erlangen Centre for Astroparticle Physics, Nikolaus-Fiebiger-Str. 2, Erlangen 91058, Germany
}

\date{Accepted XXX. Received YYY; in original form ZZZ}

\pubyear{2023}

\begin{document}
\label{firstpage}
\pagerange{\pageref{firstpage}--\pageref{lastpage}}
\maketitle

\begin{abstract}

Probabilistic classification of unassociated \Fermi-LAT sources using machine learning methods has an implicit assumption that
the distributions of associated and unassociated sources are the same as a function of source parameters,
which is not the case for the \Fermi-LAT catalogs.
The problem of different distributions of training and testing (or target) datasets as a function of input features (covariates)
is known as the covariate shift.
In this paper, we, for the first time, quantitatively estimate the effect of the covariate shift on the multi-class classification
of \Fermi-LAT sources.
We introduce sample weights proportional to the ratio of unassociated to associated source
probability density functions so that associated sources in areas, which are densely populated with unassociated sources, have more weight than
the sources in areas with few unassociated sources.
We find that the covariate shift has relatively little effect on the predicted probabilities, i.e., the training can be performed either with weighted or with unweighted 
samples, which is generally expected for the covariate shift problems.
The main effect of the covariate shift is on the estimated performance of the classification.
Depending on the class, the covariate shift can lead up to 10 -- 20\% reduction in precision and recall compared to the estimates, where the covariate shift is not taken into account.

\end{abstract}





\section{Introduction}


Classification of unassociated \Fermi-LAT sources with machine learning (ML) provides an opportunity 
to probabilistically determine the classes of sources based on their gamma-ray properties,
when direct multi-wavelength association is not known
\citep{2012ApJ...753...83A, 2016ApJ...820....8S, 2016ApJ...825...69M, 2017A&A...602A..86L, 2020MNRAS.492.5377L, 
2021RAA....21...15Z, 2021MNRAS.507.4061F, 2022A&A...660A..87B, 2023MNRAS.521.6195M}.
For some of the unassociated sources it may even be impossible to detect an associated source,
e.g., for pulsars with a radio jet that is not pointing at the observer.
In this case, the probabilistic classification of unassociated sources is the only possibility to 
determine the likely nature of the unassociated sources and to perform population studies including the unassociated sources.

\begin{figure*}
\includegraphics[width=\onesize\columnwidth]{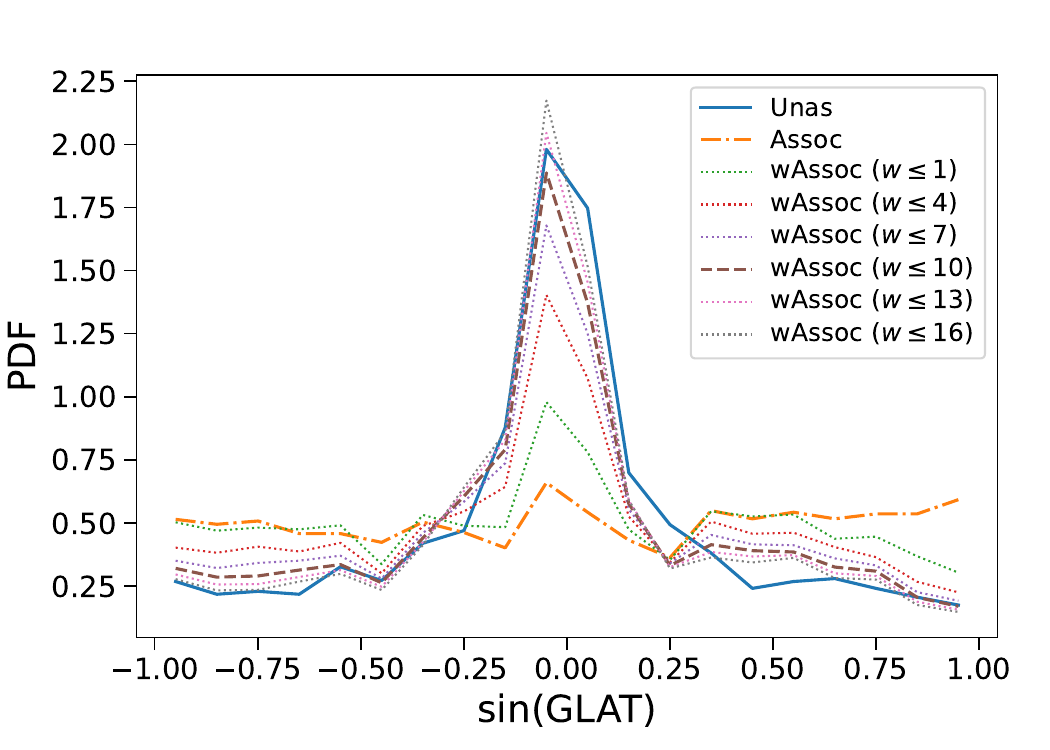}
\includegraphics[width=\onesize\columnwidth]{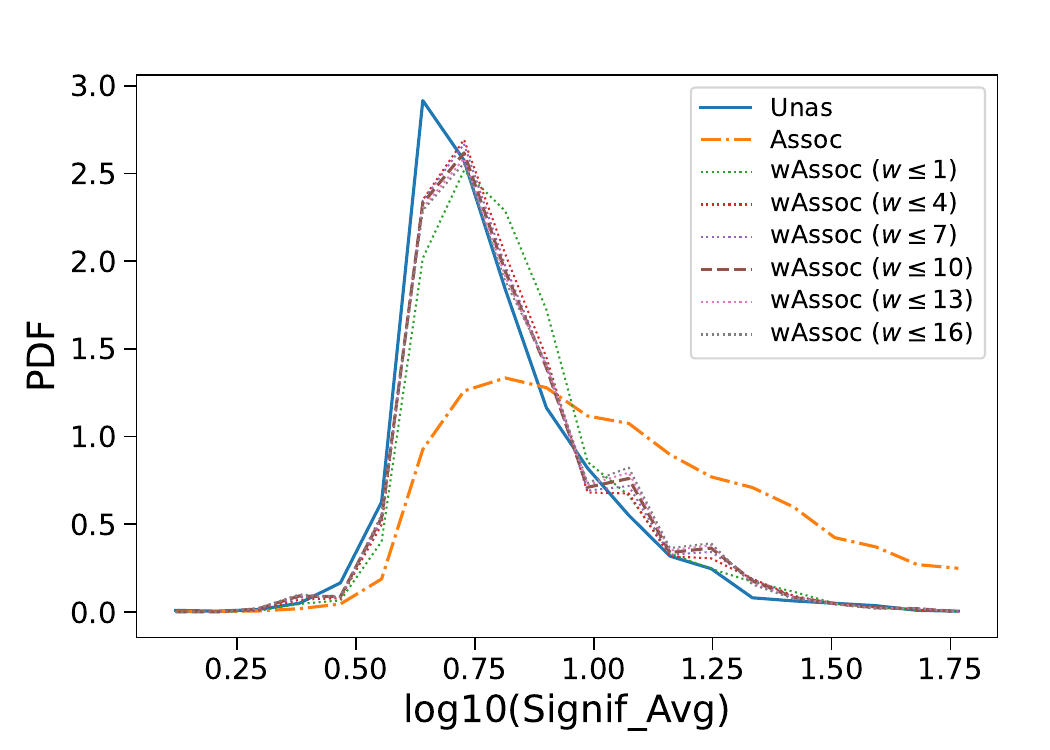} \\
\includegraphics[width=\onesize\columnwidth]{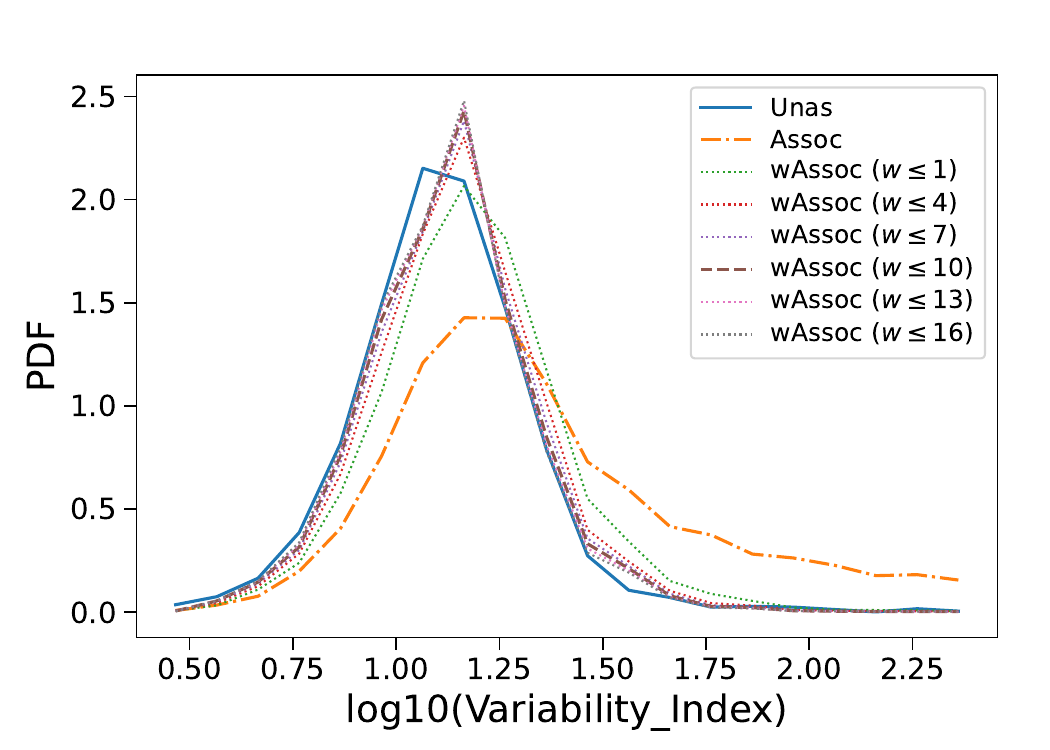}
\includegraphics[width=\onesize\columnwidth]{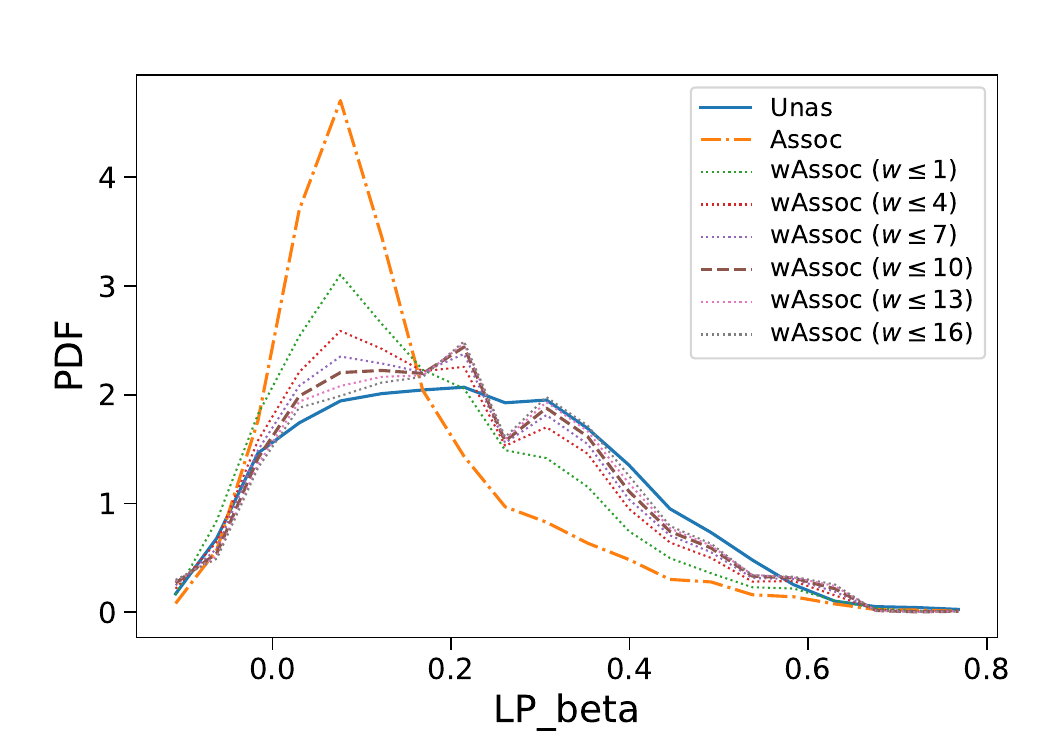}
\caption{
PDFs of unassociated (``Unas'') and associated (``Assoc'') sources as a function of the sine of Galactic latitude (upper left panel),
log of the average significance of the source (upper right panel), log of the variability index (lower left panel), and the curvature of the log parabola spectrum (lower right panel)
in the 4FGL-DR4 catalog \citep[][]{2023arXiv230712546B}.
The weighted PDFs of associated sources (``wAssoc'' labels) are obtained by multiplying the associated sources with a weighting factor proportional to the 
ratio of PDFs of unassociated to associated sources in Eq. (\ref{eq:w}). The PDFs are modeled by Gaussian mixture models (see text for more details).
The values in parentheses show the maximal sample weights.
}
\label{fig:cov_shift_examples}
\end{figure*}

One of the caveats of ML classification of \Fermi-LAT sources is that the distributions of 
associated and unassociated sources in the feature space are different.
For example, the fraction of associated sources at high latitudes is about 90\%, while the association fraction
along the Galactic plane is about 50\% \citep{2020ApJS..247...33A}.
One of the reasons is that the density of gamma-ray sources is larger along the Galactic plane (GP), 
while there is also absorption in optical and soft X-ray bands by dust and gas respectively, 
which complicates the detection of possible multi-wavelength counterparts.
In Fig.~\ref{fig:cov_shift_examples} (upper left panel),
we show the probability distribution functions (PDFs) for associated (``Assoc'' label)
and unassociated (``Unas'' label) sources
as a function of the sine of Galactic latitude (Glat).%
\footnote{
In the paper, we use the Forth \Fermi-LAT catalog data release four \citep[4FGL-DR4,][]{2023arXiv230712546B}
file version ``gll\_psc\_v32.fit''.
}
Another example is the brightness of the sources: 
brighter sources are associated more often than the dimmer ones.
The distributions of the source detection significance for associated and unassociated sources are shown in 
Fig.~\ref{fig:cov_shift_examples} upper right panel.
On the lower panels of Fig.~\ref{fig:cov_shift_examples}, we show the  
variability index  -- the significance of temporal variability, which is often used for identification of counterparts 
based on correlated flares in blazars or pulsed emission in pulsars,
and the log parabola of the beta coefficient (LP\_beta) -- the curvature of the log parabola fit of the spectrum 
(curved spectra are more typical for Galactic sources, such as pulsars).
We see that, generally, associated and unassociated sources have different distributions as a function of features.
Differences in the distribution of the training set (associated sources) and the target set (unassociated sources) 
may lead to biased predictions, such as the classes of unassociated sources,
as well as wrong estimates of the classification performance 
\citep{2020MNRAS.492.5377L, 2021RAA....21...15Z, 2021MNRAS.507.4061F, 2022A&A...660A..87B}.

The basic assumption of the ML classification is that the joint distribution of the input features $x$
and output features $y$ are the same for the training and target datasets:
\be
p_{\rm train} (x, y) = p_{\rm target} (x, y),
\ee
while, in general, a dataset shift represents a situation when the training and target distributions are different
$p_{\rm train} (x, y) \neq p_{\rm target} (x, y)$.%
\footnote{Often the target dataset is called the test dataset. 
In this paper, we use the associated sources both for training and for testing.
In order to avoid possible confusion, we use target dataset to denote unassociated sources,
which has a different distribution from both the training and testing datasets sampled from associated sources.}
The joint distribution can be written as a product of conditional probability times a prior distribution in two different ways:
\be
p(x, y) = p(y|x) p(x) = p(x|y) p(y)
\ee
Correspondingly, there are two special cases of the dataset shift \citep{MorenoTorres2012AUV}:
\ben
\item
Covariate shift: $p_{\rm train}(y|x) = p_{\rm target}(y|x)$, but $p_{\rm train}(x) \neq p_{\rm target}(x)$.
It represents the situation, when the conditional probability of a class given the input features is unchanged, but the distributions
of samples as a function of input features are different for training and target datasets.
\item
Prior shift: $p_{\rm train}(x|y) = p_{\rm target}(x|y)$, but $p_{\rm train}(y) \neq p_{\rm target}(y)$.
It represents the situation, when the prior probabilities for classes change 
(e.g., the overall fraction of sources is different for training and testing datasets),
while the distribution of input variables for each class is unchanged.
\een
In this paper, we assume that the observational limitations and association biases, which lead to the differences in the 
distributions of associated and unassociated sources affect all source classes in the same way, i.e.,
that the conditional probabilities remain the same as a function of input features: $p_{\rm train}(y|x) = p_{\rm target}(y|x)$.
In this case the dataset shift corresponds to the shift of covariates (input features): $p_{\rm train}(x) \neq p_{\rm target}(x)$.
The main goal of the paper is to determine the effect of the covariate shift on the multi-class classification of 
unassociated \Fermi-LAT sources.

An independent test of classification performance has been obtained before by cross-matching 
predictions for unassociated sources in an earlier catalog, e.g., the Third \Fermi-LAT catalog (3FGL), with the associations in the newer 4FGL catalog
\citep{2020MNRAS.492.5377L, 2021RAA....21...15Z, 2021MNRAS.507.4061F, 2022A&A...660A..87B}.
It has been observed that the performance with the cross-matching method is worse than the performance estimated 
from the testing datasets sampled from the associated sources and it was argued that this decrease in performance is due to the covariate shift
\citep{2020MNRAS.492.5377L, 2021RAA....21...15Z, 2021MNRAS.507.4061F, 2022A&A...660A..87B}.
Nevertheless, there are several issues with the cross-matching method, which we address in this paper:
\ben
\item
The sample of sources in the cross-matching dataset is not representative of the total population of unassociated sources.
\item
The reduction of performance can be partially due to uncertainties in the reconstruction of the source parameters \citep[e.g.,][]{2022A&A...660A..87B}.
However, such uncertainties do not lead to covariate shift: if the intrinsic distributions of associated and unassociated sources are the same, i.e.,
$p_{\rm train}(x) = p_{\rm target}(x)$ and the uncertainties depend only on features $x$, then the observed distributions 
of training and target datasets  remain the same.
The results of the cross-matching method depend both on the uncertainties on the features $x$ and on the covariate shift.
In this work we separate the two effects and determine the influence of only the covariate shift on the classification performance.
\item
The cross-matching method cannot be used for the latest catalog (there is no newer catalog that can be used for cross-matching).
\een

The paper is organized as follows. In Section \ref{sec:data} we introduce the data and define the classes.
We describe the model for the covariate shift in Section \ref{sec:cov_shift_model}.
In Section \ref{sec:cs_effect} we determine the effect of the covariate shift on two- and six-class classification
of the \Fermi-LAT sources.
We construct probabilistic catalogs of the \Fermi-LAT sources including the effect of the covariate shift
in Section \ref{sec:catalogs}.
The conclusions are presented in Section \ref{sec:conclusions}.
In Appendix \ref{app:GMM_Model} we discuss the models for the distributions of associated and unassociated sources in the feature space.
We discuss the selection of input features and their importance in Appendix \ref{app:features}.
In Appendix \ref{app:NNs} we provide details about the neural networks (NN) classification, whereas in the main body of the paper we use the random forest (RF) algorithm.

\section{Data selection and definition of classes}
\label{sec:data}

As input, we use the parameters, which describe the main features of the gamma-ray sources,
such as the position on the sky, energy spectrum, and temporal variability.
In particular, we use following 10 features derived from the source parameters in the 4FGL-DR4 catalog \citep{2023arXiv230712546B}
(see also description in \cite{2022ApJS..260...53A, 2023MNRAS.521.6195M}
and Appendix \ref{app:features} for details on the feature selection):
sin(GLAT), 
cos(GLON), 
sin(GLON), 
$\log_{10}$(Energy\_Flux100), 
$\log_{10}$(Unc\_Energy\_Flux100), 
$\log_{10}$(Signif\_Avg), 
LP\_beta,
LP\_SigCurv,
$\log_{10}$(Variability\_Index),
and the index of the log parabola spectrum at 1 GeV.
Although there are many more parameters in the 4FGL-DR4 catalog,
most of the parameters either describe the same quantity (such as the Galactic and equatorial coordinates of the sources)
or are highly correlated \citep{2022A&A...660A..87B}.
It has also been shown that, at least in case of the two-class classification, relatively few input features, e.g., five,
can provide an optimal classification performance \citep{2020MNRAS.492.5377L}.
In this work, we use the features similar to the features used in 
 \citep{2020MNRAS.492.5377L, 2022A&A...660A..87B}
as well as in the earlier works, e.g., \citep{2016ApJ...820....8S}, with some modifications described in Appendix~\ref{app:features}.
Four sources in the catalog have missing features: 4FGL J0358.4-5446 (nova),
4FGL J0534.5+2201i (pulsar wind nebula, PWN), 
4FGL J1820.8-2822 (nova), and 4FGL J2010.2-2523 (flat spectrum radio quasar, FSRQ).
We exclude these four sources from the analysis in this paper.

We use the labels for the classes of the gamma-ray sources from the 4FGL-DR4 catalog 
\citep{2022ApJS..260...53A, 2023arXiv230712546B}
and consider identified sources (upper-case class names in 4FGL-DR4) and associated sources (lower-case class names) on the same footing.
The physical classes of sources are summarized in Table~\ref{tab:assoc_classes}.

\begin{table}
\centering
\caption{Classes of associated sources in the 4FGL-DR4 catalog \citep{2022ApJS..260...53A, 2023arXiv230712546B}.
Both associated and identified sources in the catalog are referred as associated sources in this work.}
\label{tab:assoc_classes}
\begin{tabular}{lll}
\hline
Physical class & Assoc. sources & Description \\
\hline
gc & 1 &  Galactic center\\
psr & 141 &  young pulsar\\
msp & 179 &  millisecond pulsar\\
pwn & 21 &  pulsar wind nebula\\
snr & 45 &  supernova remnant\\
spp & 124 &   supernova remnant \\
 & &   and/or pulsar wind nebula \\
glc & 34 &  globular cluster\\
sfr & 6 &  star-forming region\\
hmb & 11 &  high-mass binary\\
lmb & 9 &  low-mass binary\\
bin & 10 &  binary\\
nov & 6 &  nova\\
bll & 1490 &  BL Lac type of blazar\\
fsrq & 819 &  FSRQ type of blazar\\
rdg & 53 &  radio galaxy\\
agn & 8 &  non-blazar active galaxy\\
ssrq & 2 &  steep spectrum radio quasar\\
css & 6 &  compact steep spectrum radio source\\
bcu & 1624 &  blazar candidate of uncertain type\\
nlsy1 & 8 &  narrow-line Seyfert 1 galaxy\\
sey & 3 &  Seyfert galaxy\\
sbg & 8 &  starburst galaxy\\
gal & 6 &  normal galaxy (or part) \\ 
\hline
\end{tabular}
\end{table}

We consider sources with unknown nature of the multi-wavelength counterpart (labeled as ``unk'' in the catalog) as unassociated sources.
Overall, we have 4614 associated, and 2577 unassociated sources.
Note that the total number of sources is 7191, which is less than the number of sources 7195 in the 4FGL-DR4 catalog \citep{2023arXiv230712546B}
by the four sources with missing input features.

\begin{figure}
\includegraphics[width=\onesize\columnwidth]{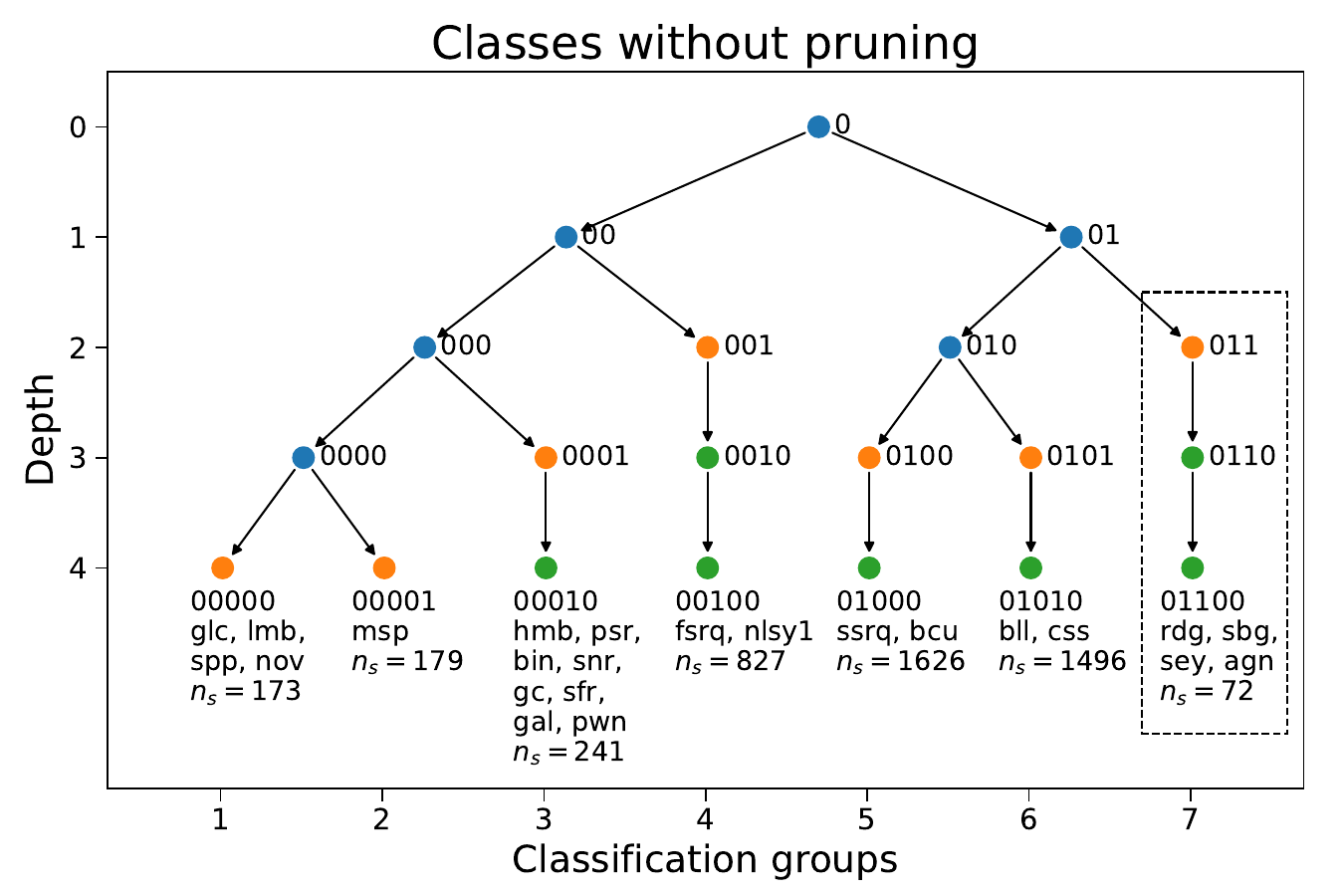}
\includegraphics[width=\onesize\columnwidth]{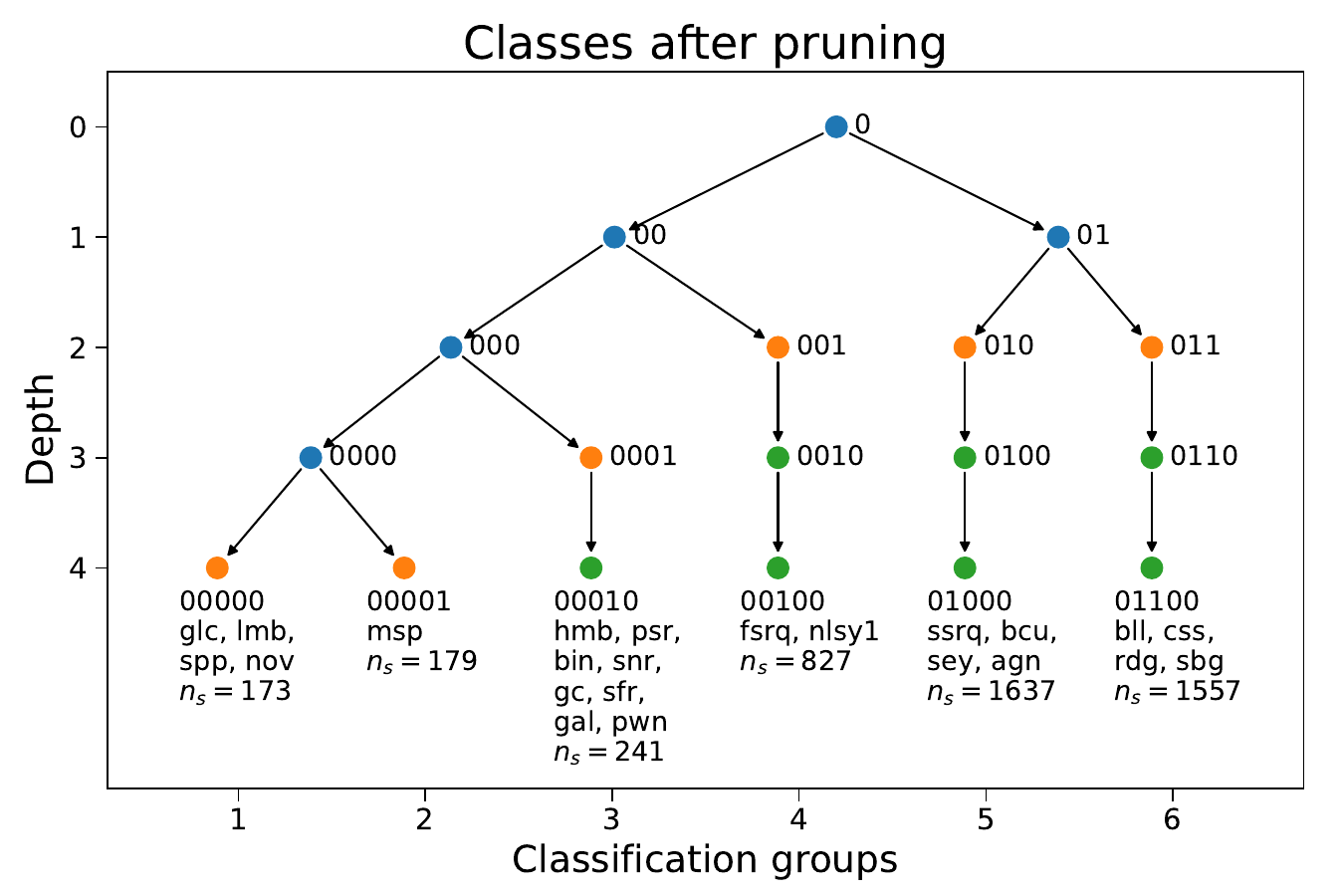}
\caption{
Definition of classes. Top panel: hierarchical definition of classes \citep[following the method of][]{2023MNRAS.521.6195M}
with the condition that the number of sources in a class is larger than 50.
Dashed box shows the class with the smallest number of sources. We remove this class from the final definition of classes. The corresponding physical classes 
(rdg, sey, sbg, and agn) are distributed among the remaining six classes according to the maximal class probability sum for each of the physical class.
Since node 01 has only one child 010 after pruning, we merge the nodes 01 and 010 into a new node 01.
Bottom panel: the result of pruning, which shows the final hierarchical structure of the classes used for the classification in this paper.
See text for more details.}
\label{fig:class_def}
\end{figure}

Provided that some of the physical classes have too few members for a reasonable classification (e.g., less than 10 associated sources),
we use an hierarchical procedure to determine the classes \citep{2023MNRAS.521.6195M}
that combines several physical classes with similar distributions in the feature space.
In particular, we use the Gaussian mixture model (GMM) to determine the hierarchical splitting of the physical classes \citep[for details, see][]{2023MNRAS.521.6195M}.
An example of such splitting of the physical classes with the condition on the minimal number of sources in a class $n_{\rm s} > 50$ is shown
in Fig.~\ref{fig:class_def}, top panel.
We note that node 011, which includes rdg, sey, sbg, and agn classes, has only 74 associated sources.
We have checked that the classification into seven classes corresponding to the terminal nodes in the top panel of Fig.~\ref{fig:class_def}
does not give reasonable results for the 011 class.
However, if we increase the condition on the minimal number of sources to be, e.g., $n_{\rm s} > 100$, then 
node 01 cannot split.
As a result, this node has 3194 sources, which is almost 70\% of all associated sources.
In this case, the multi-class classification is not meaningful either.
A possible solution to this problem is to first construct the classes with the condition $n_{\rm s} > 50$ and then prune the tree by removing the node with the 
minimal number of classes, e.g., node 011.
The removed node is shown by the dashed box in the top panel of Fig.~\ref{fig:class_def}.
Since the parent node 01 has now only one child node 010, 
we merge nodes 010 and 01, i.e., the subtree under 010 is now a subtree under 01 (the corresponding nodes in the subtree move one level up).
The resulting tree is shown in Fig.~\ref{fig:class_def}, bottom panel.

The physical classes in the pruned node are distributed among the remaining six classes.
In order to determine that, we train RF classification with the six classes and then classify sources in  
rdg, sey, sbg, and agn classes using the six-class classification.%
\footnote{In this paper we use RF with maximal number of trees 50 and maximal depth of 15. 
For the other parameters we use default values in the scikit-learn version 1.2.2 \citep{scikit-learn} implementation of RF.
In particular, the Gini index is used for the determination of the splits.}
We compute the sum of class probabilities for all sources in each of the rdg, sey, sbg, and agn classes
and attach these classes to groups with the largest sum of class probabilities.
The result is that 
the sey and agn classes are attributed to node 010 dominated by the bcu class, while the rdg and sbg classes are attributed to node 011 dominated by the bll class.
The result of this procedure is shown in the bottom panel of Fig.~\ref{fig:class_def}.
We also show the summary of the remaining six classes in Table~\ref{tab:classes}.

\begin{table}
\centering
\caption{Definition of classes.
The classes are labeled by the largest physical class, e.g., spp+ or msp+.}
\label{tab:classes}
\begin{tabular}{llc}
\hline
Class label & Physical classes & Assoc. sources \\
\hline
spp+ & glc, lmb, spp, nov & 173 \\
msp+ & msp & 179 \\
psr+ & hmb, psr, bin, snr, & \\
  & gc, sfr, gal, pwn & 241 \\
fsrq+ & nlsy1, fsrq & 827 \\
bcu+ & ssrq, bcu, sey, agn & 1637 \\
bll+ & bll, css, rdg, sbg & 1557 \\
\hline
\end{tabular}
\end{table}

We note that physical classes are grouped in the six groups according to their gamma-ray properties, 
i.e., even if the physical nature of the sources is different but the gamma-ray properties are similar, 
the physical classes would be added to the group.
For example, the ``bll+'' group has mostly active galactic nuclei (bll, css, and rdg) and the starburst galaxies class (sbg).
A comparison of the two most important features for the separation of the physical classes at level one 
(``LP\_beta'' and ``log10(Unc\_Energy\_Flux100)'')
for bll, sbg and several other physical classes can be found in Figure 1 of \cite{2023MNRAS.521.6195M}.
The assembly of the groups according to the similarities in the gamma-ray properties ensures an optimal multi-class classification performance.
This is a fundamental limitation of any ML classification that the gamma-ray properties used for the classification do not necessarily reflect the different physical nature of the sources.
In particular, in our analysis the ML classification cannot separate starburst galaxies from other sources in the bll+ class.
Although the multi-class classification is dominated by the large classes, such as spp, msp, psr, fsrq, bcu, and bll,
and, as a result, it is also mostly useful for the separation of these large classes, 
it nevertheless, can be useful also for the small classes,
as it can provide additional evidence for association or classification.
For example, if there is a nearby starburst galaxy in a vicinity of an unassociated source with high bll+ classification probability, then the source is more likely to be a starburst galaxy compared to the situation, when this source is classified as a  member of, e.g., msp+ or fsrq+ classes.


\section{Covariate shift model}
\label{sec:cov_shift_model}

\begin{figure}
\includegraphics[width=\onesize\columnwidth]{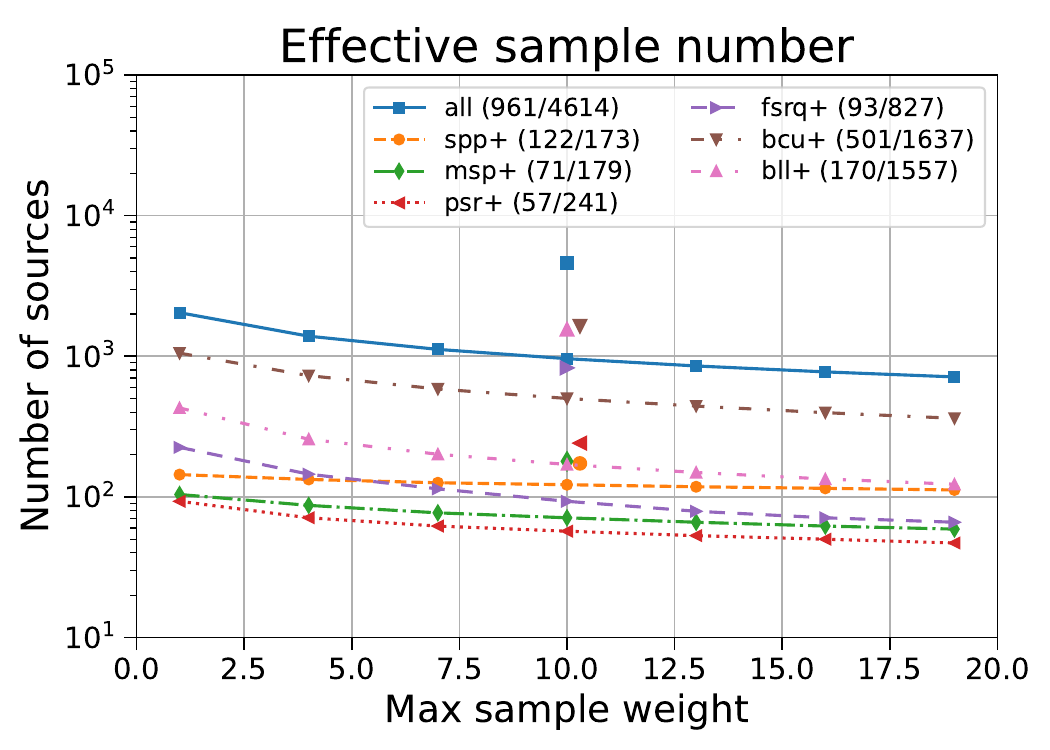}
\includegraphics[width=\onesize\columnwidth]{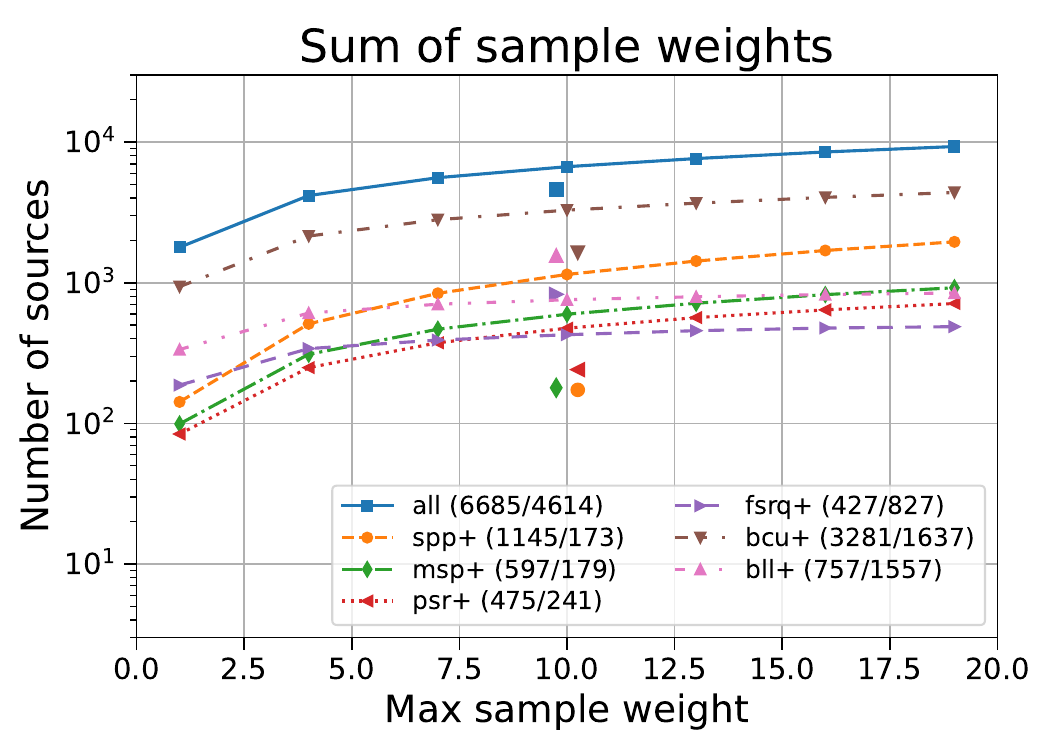}
\caption{
Effective number of samples (top) and oversampling (bottom) as a function of the maximal sample weight.
Top panel: effective number of samples as defined in Eq. (\ref{eq:neff}).
The numbers in parentheses show respectively the effective number of samples at $w_{\rm max} = 10$ and the total number of 
associated sources in each of the six classes (Table \ref{tab:classes}).
The corresponding numbers of associated sources are also shown as the stand-alone points near $w_{\rm max} = 10$.
Bottom panel: oversampling (or undersampling) of sources defined by summing the sample weights.
The first number in parentheses shows the oversampled number of sources for all associated sources and for each of the six classes
at $w_{\rm max} = 10$.
The second number is the number of associated sources (also shown as stand-alone points).
}
\label{fig:sample_weight}
\end{figure}

The presence of the covariate shift manifests itself in the fact that the ratio of the training and the target PDFs is not a constant 
in the multi-dimensional feature space.
Provided that the domains of the training and the target datasets are the same for the associated and unassociated sources,
the effect of the covariate shift can be modeled by introducing weights for samples in the training and testing dataset 
proportional to the ratio of the corresponding PDFs
\be
\label{eq:w}
w(x_i) = \frac{p_{unas}(x_i)}{p_{assoc}(x_i)}.
\ee
In this case the differences in the densities of training or testing and target datasets is compensated by the weighting of the samples.
In order to determine the weighting factor $w(x)$ as a function of the features,
one needs to model the PDFs $p_{\rm unas}(x_i)$ and $p_{\rm assoc}(x_i)$.
There are different ways to approximate a distribution of discrete points with a continuous PDF, e.g., using kernel density estimators.
In this paper, we use GMMs for modeling the PDFs of associated and unassociated sources.
Details about the construction of the PDF models are provided in Appendix \ref{app:GMM_Model}.
In order not to give too much weight to any of the sources, we put an upper bound on the weights.
Examples of the PDFs for the associated sources including sample weights are presented in Fig. \ref{fig:cov_shift_examples}
(``wAssoc'' labels) with several values of the maximal weight, e.g., $w \leq w_{\rm max} = $ 1, 4, \ldots, 16.
Larger maximal weights typically give a better agreement between the distribution of unassociated sources and the weighted associated sources, especially for the Galactic latitude distribution.
However, for most of the features the dependence on $w_{\rm max}$ is not very significant.
Also larger maximal weight reduces the effective number of samples, 
where for a set of samples with weights $w_i$, the effective number of samples is computed as \citep{kish1965survey}:
\be
\label{eq:neff}
n_{\rm eff} = \frac{(\sum_i w_i)^2}{\sum_i w_i^2}.
\ee
We show the effective sample number as a function of $w_{\rm max}$ in the top panel of Fig. \ref{fig:sample_weight}.
For example, the effective number of associated sources for $w_{\rm max} = 10$ is 961, which is more than four times smaller than the total number of associated sources in the 4FGL-DR4 catalog (4614), i.e., the variance of model parameters in the 
weighted sample case can be expected to be larger than in the unweighted case.

The weighting affects different classes unequally (see Table \ref{tab:classes} for the class definitions).
In particular, the effective number of fsrq+ sources is about nine times smaller than the number of associated fsrq+ sources,
while the effective number of spp+ sources is less than 1.5 times smaller than the number of associated spp+ sources.
Another characteristic number is oversampling (or undersampling), which is computed as the sum of the weights.
The sum of weights for all sources and for the six classes are shown 
in the bottom panel of Fig. \ref{fig:sample_weight}.
We see that bll+ and fsrq+ classes are undersampled by a factor of about 2,
while all other classes (including bcu+) are oversampled with an oversampling factor of up to 6.6 (for the spp+ class).
Overall, we find that $w_{\rm max} = 10$ provides a reasonable compromise between the approximation
of the distribution of the unassociated sources (Fig. \ref{fig:cov_shift_examples}) and the effective number of samples (top panel of Fig. \ref{fig:sample_weight}).
We  use the $w_{\rm max} = 10$ case below for training and testing with weighted samples.


\section{Effect of covariate shift on classification}
\label{sec:cs_effect}

\subsection{Two-class classification}


In this section,
in order to get an intuition about the effect of the covariate shift, 
we use a two-class (rather than a multi-class) classification problem.
We define the two classes as active galactic nuclei (AGNs) and Galactic sources (including other galaxies):\\
AGNs: bll, fsrq, rdg, agn, ssrq, css, bcu, nlsy1, sey (4013 sources); \\
Galactic: psr, msp, gc, pwn, snr, spp, glc, sfr, hmb, lmb, bin, nov, sbg, gal (601 sources).

\begin{figure*}
\includegraphics[width=\twosize\textwidth]{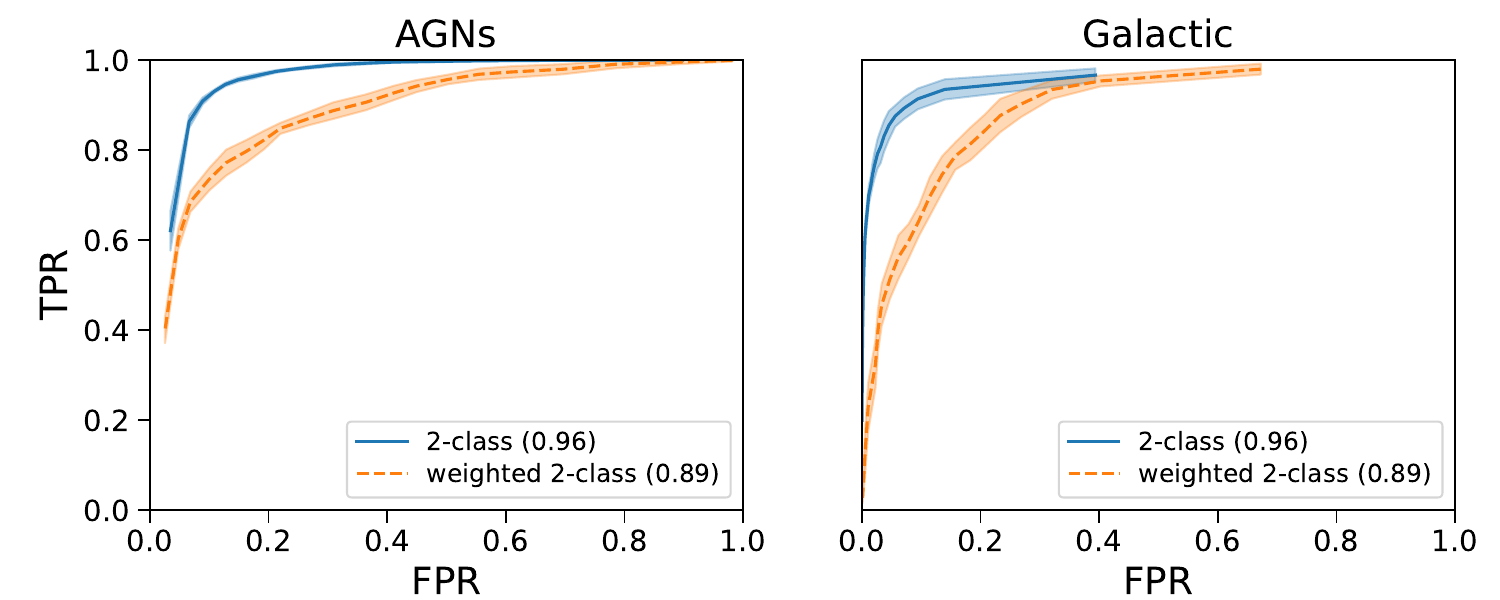}
\caption{
ROC curves for unweighted training and testing (``2-class'' labels) and weighted training and testing (``weighted 2-class'' labels).
The lines (shaded areas) represent the mean (the standard deviation) over ten splits into training and testing datasets with the 70/30\% ratio.
}
\label{fig:2class_ROC}
\end{figure*}

The classification is performed with the RF algorithm.
A comparison of receiver operating characteristic (ROC) curves for the unweighted two-class classification and for the classifications where
weights are applied both for training and testing is presented in Fig.~\ref{fig:2class_ROC}.
We see that the area under the curve (AUC) is reduced from about 0.96 to 0.89 both for AGNs and for Galactic sources.

\begin{figure*}
\includegraphics[width=\twosize\textwidth]{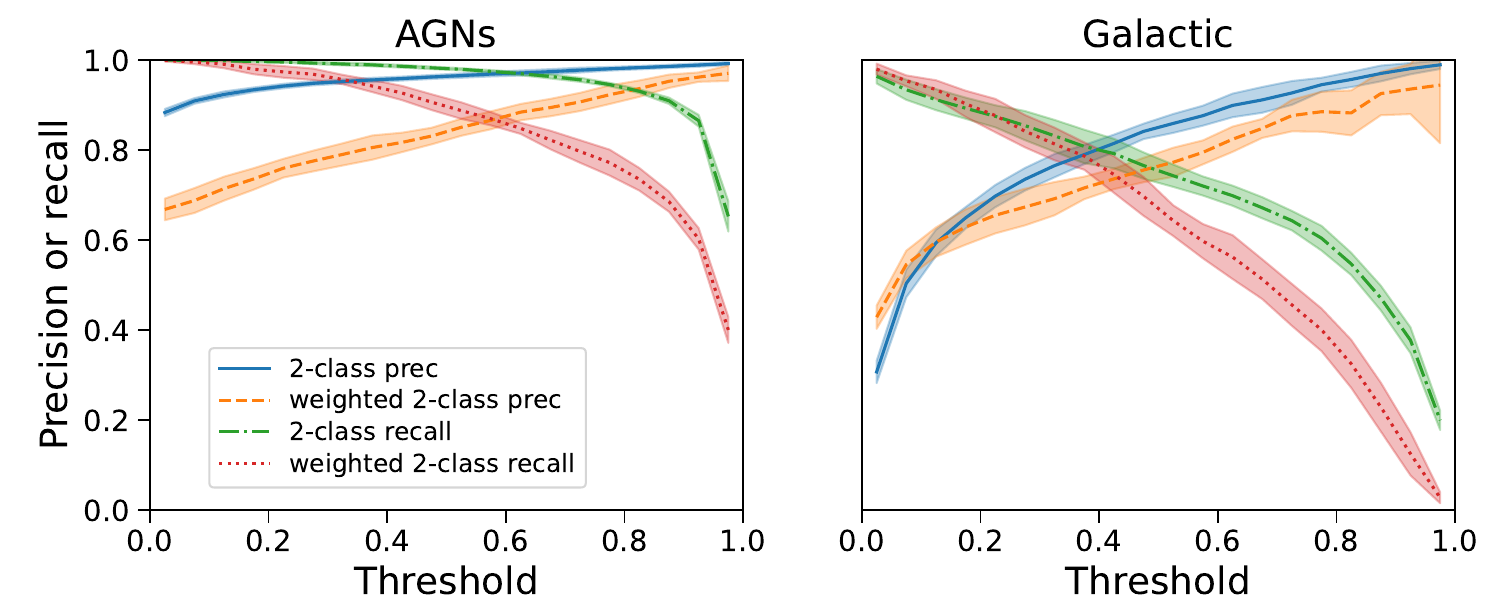}
\caption{
Comparison of precision and recall for unweighted training and testing (``2-class'' labels) and weighted training and testing (``weighted 2-class'' labels). 
The lines (shaded areas) represent the mean (the standard deviation) over ten splits into training and testing datasets with the 70/30\% ratio.
}
\label{fig:2class_PR}
\end{figure*}

The corresponding comparison of precision and recall is shown in Fig.~\ref{fig:2class_PR}.
In this case AGNs are affected slightly more than the Galactic sources relative to statistical uncertainty.

\begin{figure*}
\includegraphics[width=\twosize\textwidth]{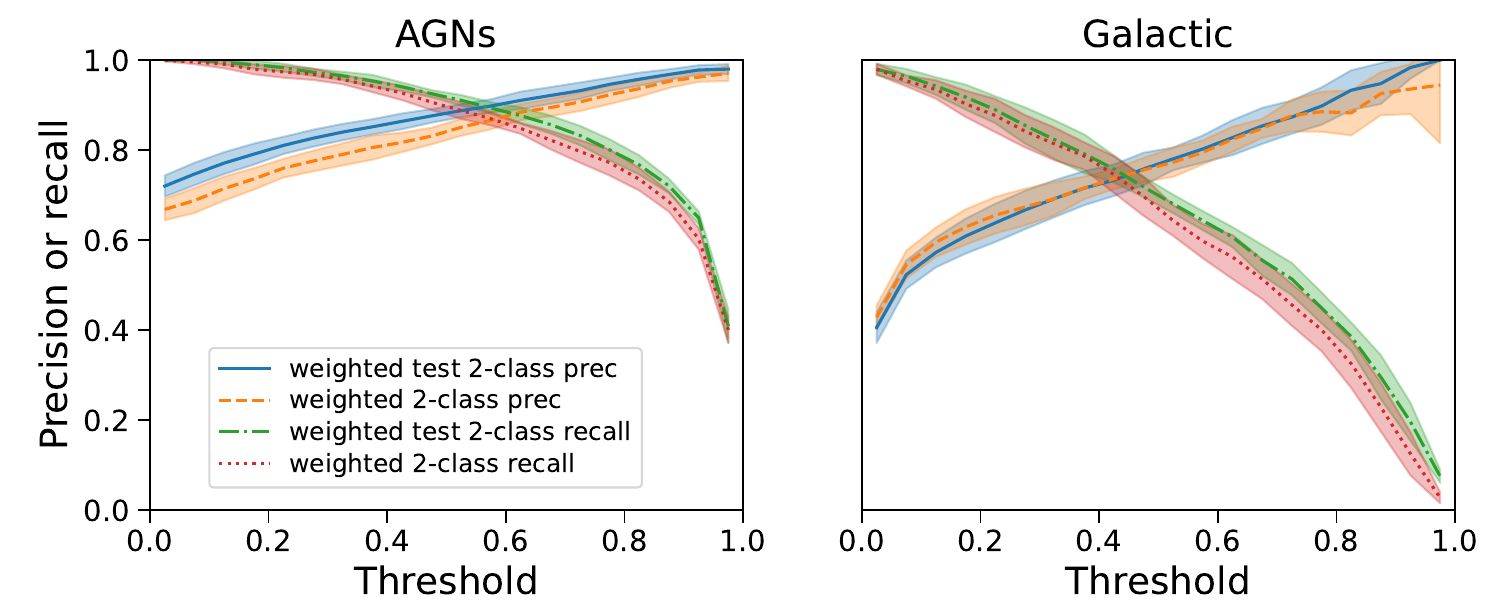}
\caption{
Comparison of precision and recall for 
unweighted training with weighted testing (``weighted test 2-class'' labels)
and weighted training and testing (``weighted 2-class'' labels).
The performance calculated on weighted test samples for training with unweighted samples
is similar to the performance of training with weighted samples.
See Fig.~\ref{fig:2class_PR} for the definition of the lines.
}
\label{fig:2class_PRwt}
\end{figure*}

It is interesting to note that if we use weighting for testing only and perform training with unweighted samples, then the performance is similar to the case
when the weighting is applied both for training and for testing.
We compare the corresponding precision and recall in Fig.~\ref{fig:2class_PRwt}.
This result is not surprising, provided that the classification algorithm learns the conditional probabilities $p(y|x)$, 
which are not affected by the weights.
Nevertheless, it does serve as a cross-check of the procedure, which shows that training with either weighted or unweighted samples can be used for the 
classification of unassociated sources.
But it is important to use weights for the testing samples in the estimation of the performance to make sure that it is estimated for the 
sources with a distribution similar to the distribution of the target dataset, i.e., the unassociated sources.

\subsection{Multi-class classification}




\begin{figure*}
\includegraphics[width=\onesize\textwidth]{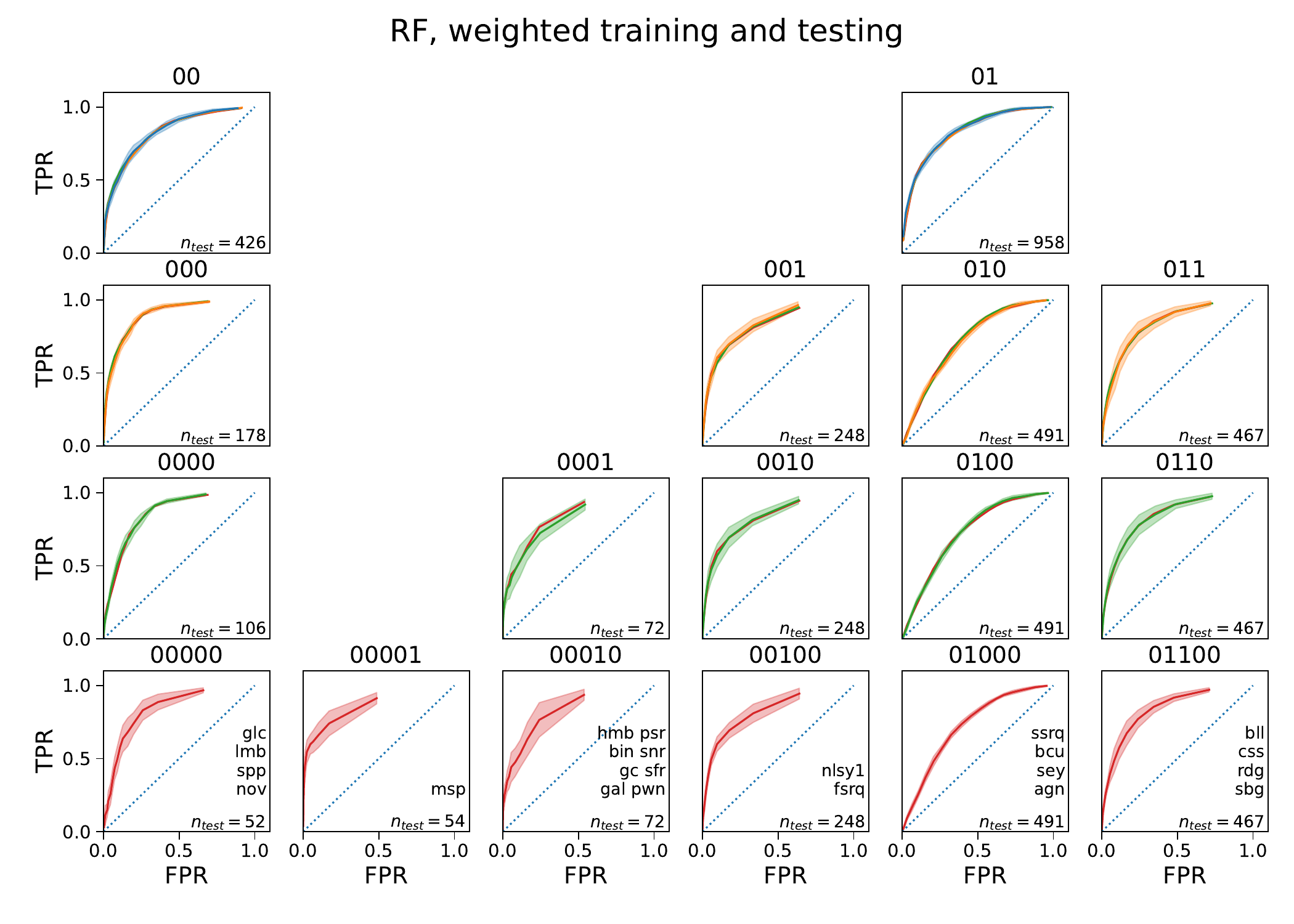}
\caption{
ROC curves for weighted training and testing following
the hierarchical definition of classes in Fig.~\ref{fig:class_def}.
The physical classes in a parent node are obtained by removing the last digits in the node names of the children nodes,
e.g., the 0000 node contains physical classes of 00000 and 00001 nodes.
At each level, the class probabilities are computed either directly for two-, four-, five-, or six-class classification or by summing the probabilities
of the children nodes.
Lines (shaded areas) show the mean (standard deviation) for 10 random splits into training and testing sets with 70/30\% ratio.}
\label{fig:mclass_ROCw}
\end{figure*}

In this section, we study the effect of the covariate shift for training and testing in multi-class classification of the \Fermi-LAT sources.
For the classification, we use the six classes summarized in Table~\ref{tab:classes} and Fig.~\ref{fig:class_def}.
As an example, we perform the classification with the RF algorithm in this section,
while in Appendix \ref{app:NNs} we compare the results with the classification using NN implemented with TensorFlow
\citep{tensorflow2015-whitepaper}.
We use 70/30\% split into training and testing datasets.
The performance is evaluated on the testing datasets. 

The ROC curves calculated using the one-vs-all definition of the true positive and the false positive rates (TPR and FPR respectively)
are shown in the bottom panels in Fig.~\ref{fig:mclass_ROCw}.
Both the training and the testing is performed with weighted samples, where the weights are determined in Eq. (\ref{eq:w}).
The shaded areas show 
the standard deviation of the ROC curves calculated from 10 random split into training and testing datasets.
The second (from the bottom) row of panels in Fig.~\ref{fig:mclass_ROCw} shows the ROC curves in the five-class classification,
where the five classes are obtained by merging some of the six classes by removing the last digit in the class names
for the 6 classes (shown in the titles of the panels). In particular, the physical classes corresponding to the 0000 node are obtained by joining the classes
in 00000 and 00001 nodes. The physical classes in the 0000 node are glc, lmb, spp, nov (which come from node 00000) and msp (which come from node 00001).
The green curves show the ROC curves in the five-class classification.
Analogously to the six-class classification, the green shaded area shows the uncertainty due to the random splits into training and testing datasets.
The red curves in these panels show the ROC curves for probabilities obtained by summing the six-class probabilities of the children nodes.
The performance for the direct five-class classification and for the five-class probabilities determined by summation of the six-class probabilities are practically the same.
This conclusion also holds for the two- and four-class classification shown in rows one and two of Fig.~\ref{fig:mclass_ROCw},
where we show both the ROC curves for the direct two- and four-class classifications and for probabilities determined by summation of class probabilities of the children nodes.
Similarly to the conclusions of \cite{2023MNRAS.521.6195M}, we find that the performance of direct classification with two, four, and five classes
and the performance of classification obtained by summation of probabilities of children nodes in the six-class case are very similar
also for the weighted samples.
Consequently, it is sufficient to do a classification with the maximal number of classes (six classes in this case).

\begin{figure*}
\includegraphics[width=\threesize\textwidth]{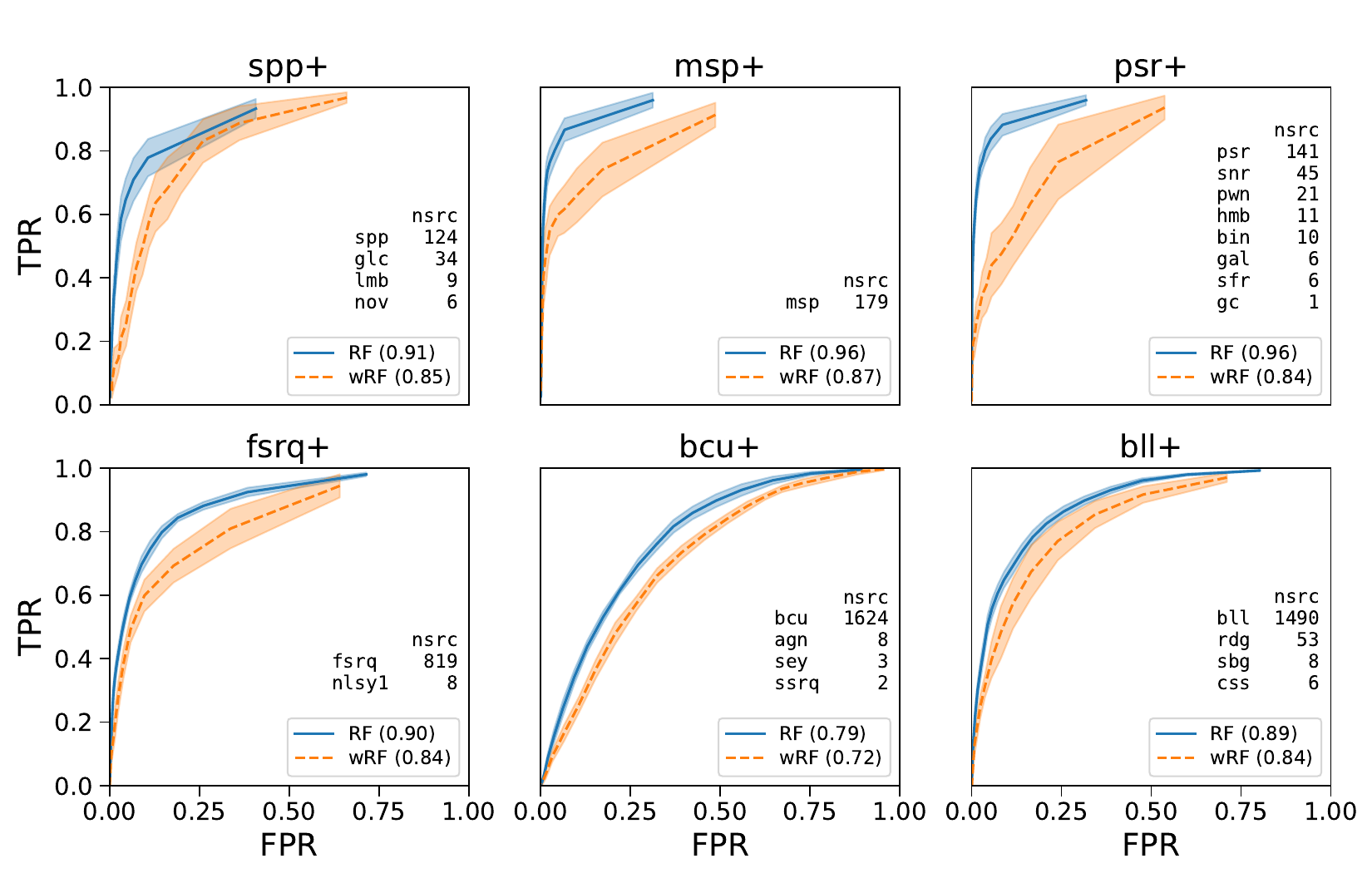}
\caption{
ROC curves for unweighted training and testing using the RF classification algorithm (``RF'' labels) and weighted training and testing (``wRF'' labels)
for the six-class classification of the \Fermi-LAT sources.
The physical classes in each group and the numbers of associated sources in each physical class are shown in tables inside the panels.
}
\label{fig:mclass_ROC_compare}
\end{figure*}

In Fig.~\ref{fig:mclass_ROC_compare} we compare the ROC curves for the weighted and unweighted multi-class classifications.
Similarly to the two-class case, the performance in the unweighted dataset is better than for the weighted one.
Provided that the unweighted dataset represents the associated sources, while the weighted one models the distribution of the unassociated sources,
the performance of the classification for the unassociated sources determined from the associated sources (without weighting)
is overestimated.

\begin{figure*}
\includegraphics[width=\threesize\textwidth]{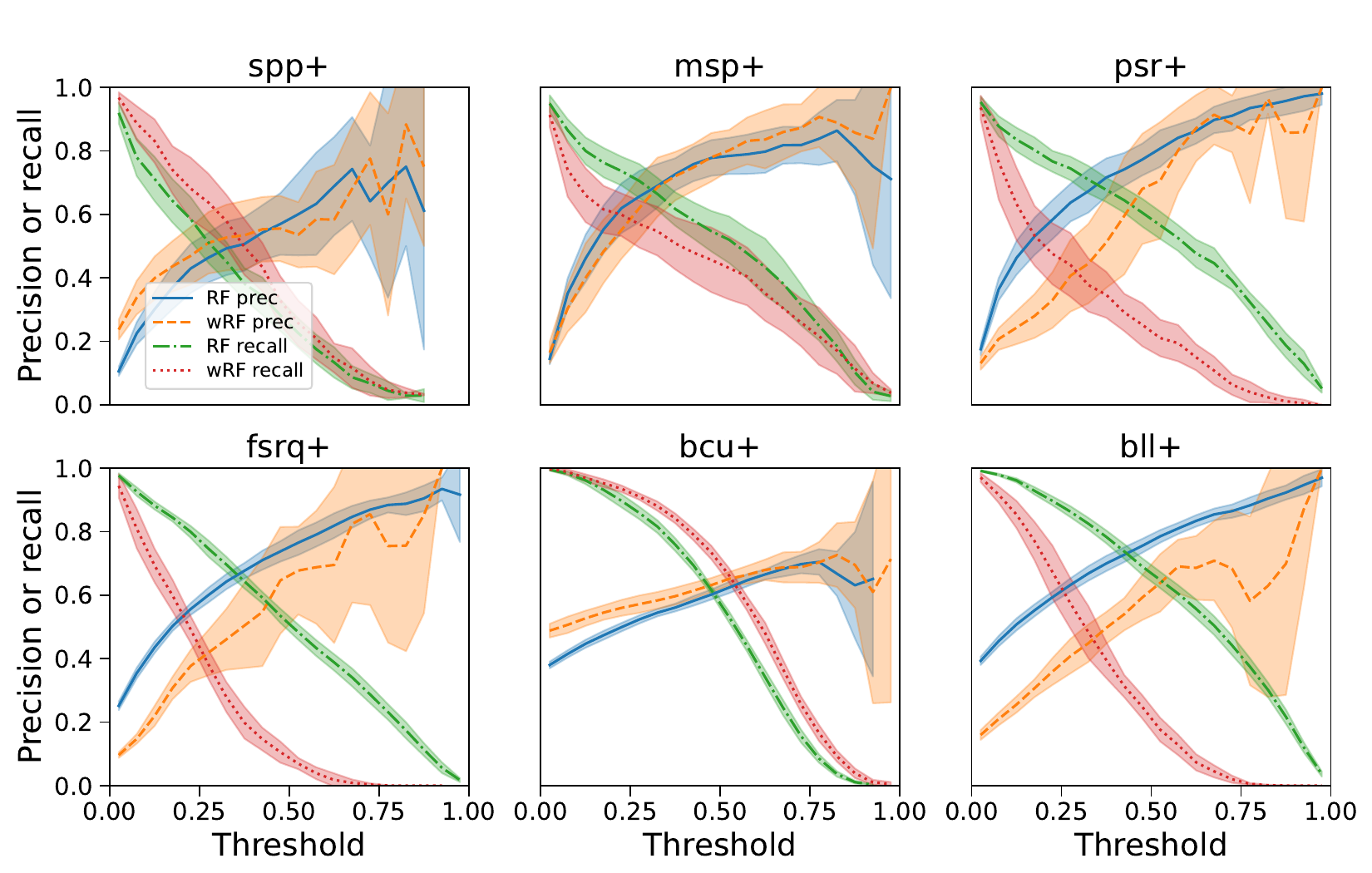}
\caption{
Precision and recall for unweighted training and testing using the RF classification algorithm (``RF'' labels) and weighted training and testing (``wRF'' labels)
for the six-class classification of the \Fermi-LAT sources. For the definition of the classes, see Fig.~\ref{fig:mclass_ROC_compare}.
}
\label{fig:mclass_PR_compare}
\end{figure*}

We show the difference in precision and recall for weighted and unweighted training and testing
in Fig.~\ref{fig:mclass_PR_compare}.
For example, the precision and recall for the bll+ and bcu+ classes in the weighted case are worse, respectively better, than in the unweighted case, 
while the effect of using the weighted samples on the ROC curves for the bll+ and bcu+ classes are comparable, i.e., 
the AUC is smaller for both classes for the weighted relative to the unweighted cases.
The worse ROC curves for the bll+ class is explained by the worse true positive rate, i.e., the recall, while the reduction of the 
ROC curve performance for the bcu+ class is explained by the worse false positive rate, which is the fraction of non-bcu+ sources
attributed to the bcu+ class. We note that the bll+ class is undersampled by about a factor of two, while the bcu+ class is oversampled by almost a factor of two (bottom panel of Fig. \ref{fig:sample_weight}).
Another example of a slightly better precision and recall in the weighted case but a worse ROC curve is provided by the spp+ class.
This can be explained by a similar increase in false positive and true positive detections but a smaller size of the ``negative'' dataset due to large oversampling of the spp+ class by a factor of 6.6
(bottom panel of Fig. \ref{fig:sample_weight}).

\begin{figure*}
\includegraphics[width=\threesize\textwidth]{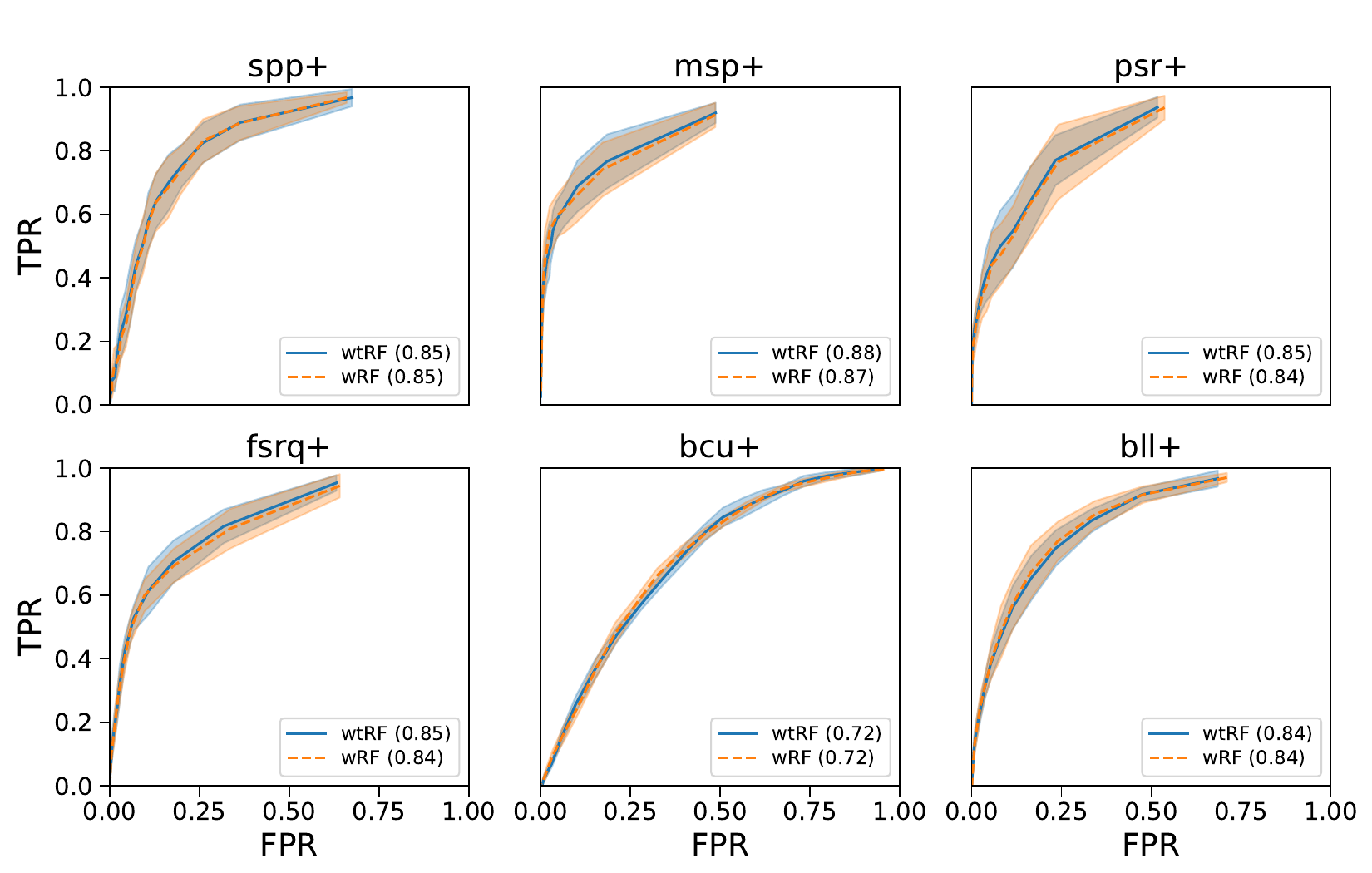}
\caption{
ROC curves for unweighted training but weighted testing using the RF classification algorithm (``wtRF'' labels) and weighted training and testing (``wRF'' labels)
for the six-class classification of the \Fermi-LAT sources.
}
\label{fig:mclass_ROC_compare_wt}
\end{figure*}

\begin{figure*}
\includegraphics[width=\threesize\textwidth]{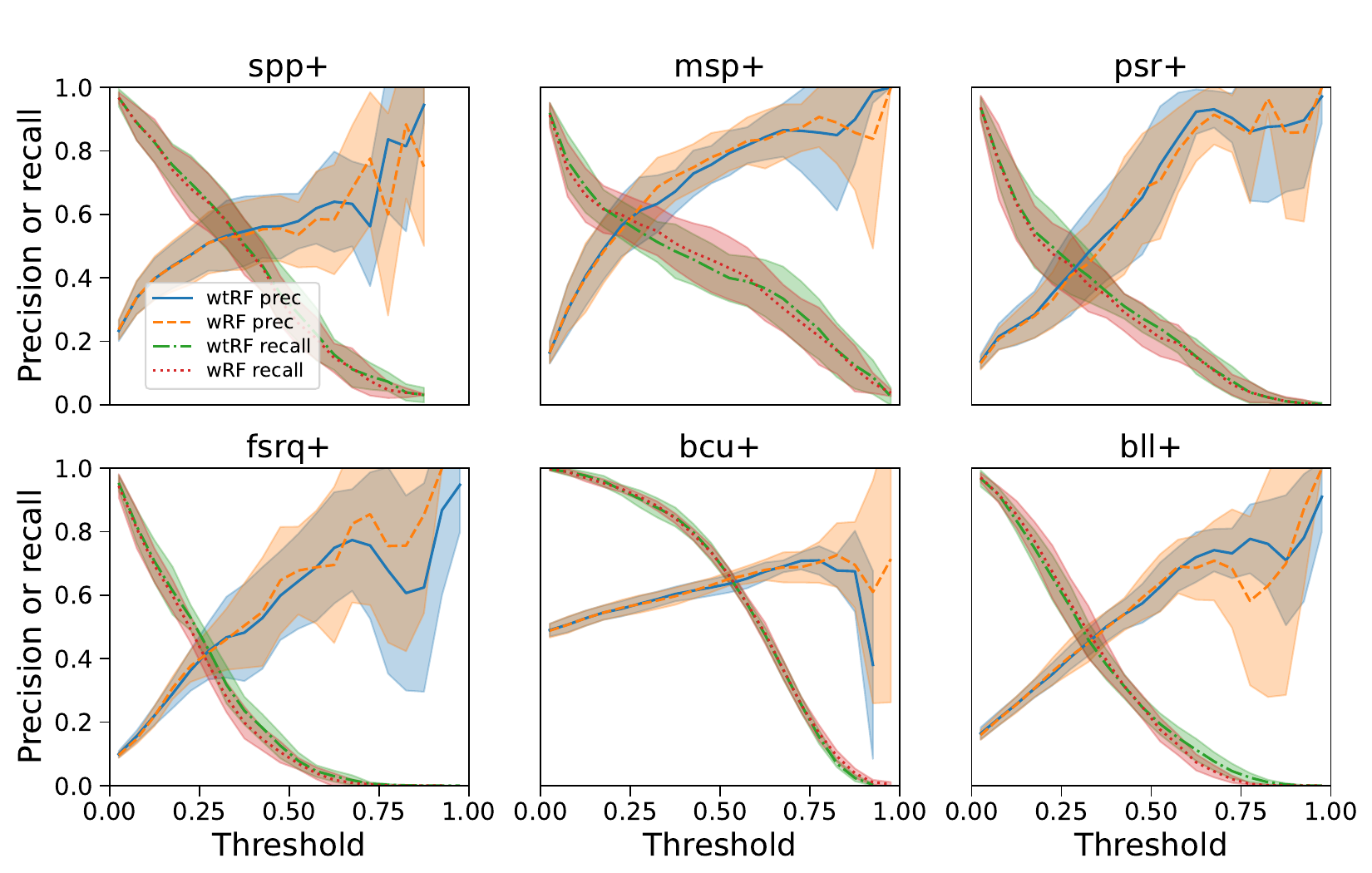}
\caption{
Precision and recall for unweighted training but weighted testing using the RF classification algorithm (``wtRF'' labels) and weighted training and testing (``wRF'' labels)
for the six-class classification of the \Fermi-LAT sources.
}
\label{fig:mclass_PR_compare_wt}
\end{figure*}

\begin{figure*}
\includegraphics[width=\threesize\textwidth]{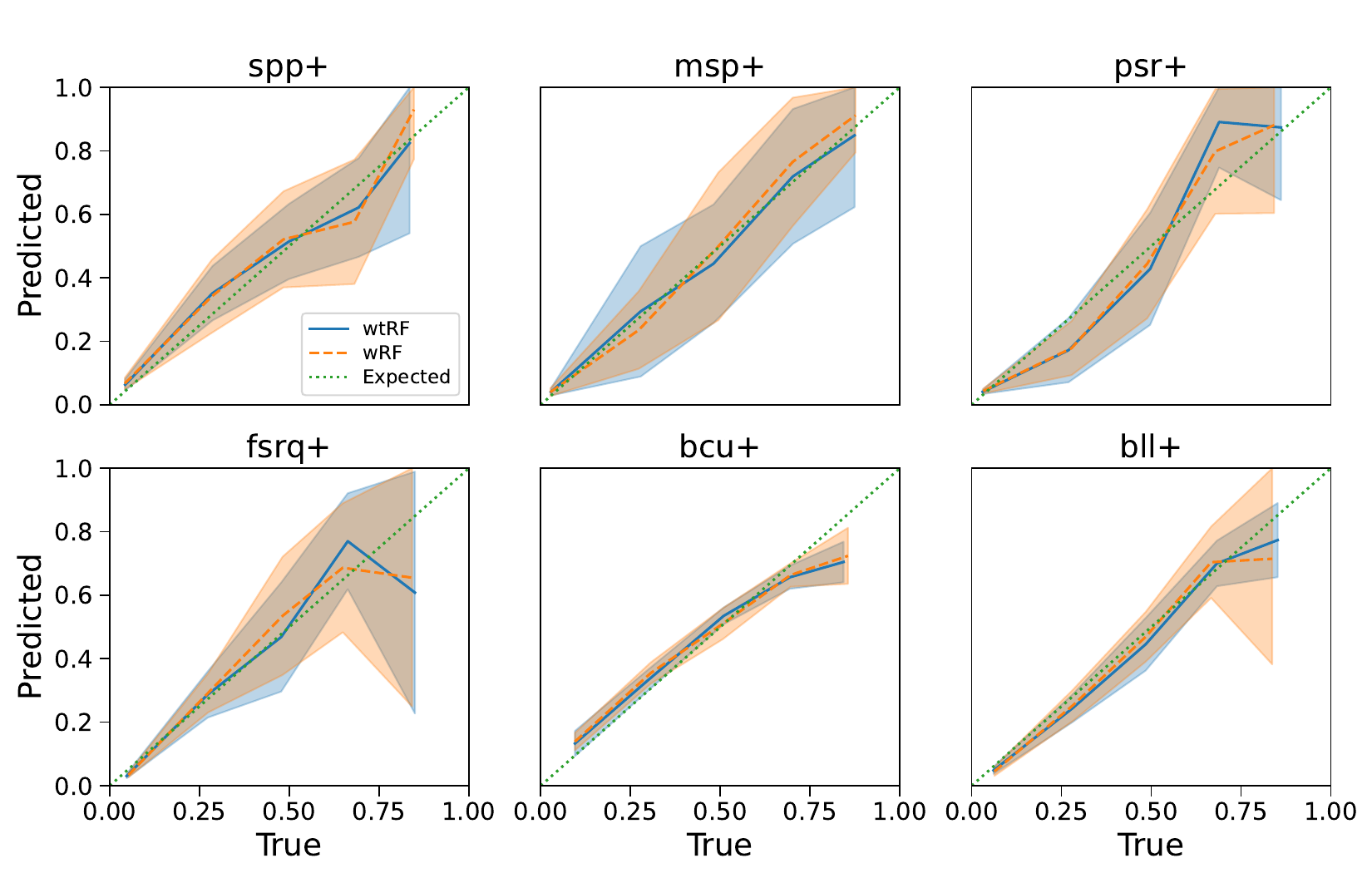}
\caption{
Reliability diagrams for unweighted training but weighted testing using the RF classification algorithm (``wtRF'' labels) and weighted training and testing (``wRF'' labels)
for the six-class classification of the \Fermi-LAT sources.
The dotted line shows the optimal calibration of the predicted probabilities.
}
\label{fig:mclass_RF_reliability}
\end{figure*}

In Figs. \ref{fig:mclass_ROC_compare_wt} and \ref{fig:mclass_PR_compare_wt} we compare the ROC curves and precision and recall for the classification
trained on unweighted samples but tested on the weighted samples (``wtRF'' labels) with the classification where both training and testing were performed on the weighted samples (``wRF'' labels).
We find a similar performance for weighted and unweighted training estimated from the tests on weighted samples, which is generally expected for the covariate shift.
We also find that the probabilities are generally well calibrated both for weighted and unweighted training when tested with the weighted samples
(Fig.~\ref{fig:mclass_RF_reliability}).

\section{Catalog construction with covariate shift}
\label{sec:catalogs}

In this section we describe the construction of probabilistic catalogs where the class probabilities are calculated using both weighted and unweighted training samples.
As in the previous section, we use six classes and ten input features.
For classification, we use RF and NN algorithms.
In order to estimate the uncertainty of prediction due to the random choice of the training samples, we perform several 70/30\% splits into training and testing datasets.
The predicted class probabilities for unassociated sources are computed as the mean over all predictions, while for the associated sources,
the probabilities are determined by the mean over the splits where a source is included in the testing sample.
In order to have a reasonable statistics for associated sources, we require that each associated source appears in the testing dataset at least five times, 
which resulted in 51 splits into training and testing datasets.
In the catalog, we report both the average class probabilities and the standard deviation of the class probabilities due to the random training/testing splits for each source.
Thus, for each source we report six average class probabilities determined with the RF algorithm, six class probabilities determined with the NN algorithm, 
and the corresponding standard deviations (24 columns in total).
We also include a column with sample weights. For the associated sources, the sample weight is equal to
the ratio of the unassociated to associated sources PDFs with the maximal weight of 10.
The sample weight for the unassociated sources is set to one.

We perform the classification using weighted and unweighted samples for training.
The predicted numbers of associated and unassociated sources in the six classes
(calculated as the sum of class probabilities) in the weighted and unweighted training cases are shown in 
Tables \ref{tab:summary_weighted} and \ref{tab:summary_unweighted} respectively.
The uncertainties are calculated as the root mean squared (RMS) of the corresponding standard deviations.
It is interesting to note that the predicted number of sources in a class among unassociated sources is similar for the RF algorithm 
and for most of classes for the NN algorithm.
However the expected number of sources in a class for associated sources is clearly biased in the weighted training case.
Overall, we find that unweighted training provides a more reasonable result than the weighted training,
because the performance evaluated on the weighted test samples is similar for weighted and unweighted training (cf. Figs.
\ref{fig:mclass_ROC_compare_wt}, \ref{fig:mclass_PR_compare_wt}, and \ref{fig:mclass_RF_reliability}),
while the predictions for the unweighted test samples are biased in the weighted training case (Table \ref{tab:summary_weighted})
and they are not biased in the unweighted training case (Table \ref{tab:summary_unweighted}).

\begin{table*}
\centering
\caption{Predictions for the number of associated and unassociated sources in the {\bf weighted} training catalog.
For the definition of the classes, see Table \ref{tab:classes}.
}
\label{tab:summary_weighted}
\begin{tabular}{lllllll}
\hline
 & Class label & N assoc & RF assoc & NN assoc & RF unas & NN unas \\
\hline
1 & spp+ & 173 & 183.0 +- 2.1 & 237.7 +- 2.8 & 406.7 +- 3.1 & 525.0 +- 3.5 \\
2 & msp+ & 179 & 183.4 +- 1.8 & 226.7 +- 4.0 & 177.1 +- 2.0 & 200.9 +- 2.4 \\
3 & psr+ & 241 & 238.8 +- 2.0 & 399.5 +- 5.3 & 187.5 +- 2.1 & 173.7 +- 2.1 \\
4 & fsrq+ & 827 & 792.5 +- 3.5 & 663.8 +- 5.7 & 197.5 +- 2.0 & 156.6 +- 1.9 \\
5 & bcu+ & 1637 & 1733.4 +- 4.7 & 1887.8 +- 6.3 & 1270.0 +- 4.1 & 1257.3 +- 4.1 \\
6 & bll+ & 1557 & 1482.8 +- 4.2 & 1198.5 +- 6.1 & 338.1 +- 2.5 & 263.4 +- 2.3 \\
\hline
\end{tabular}
\end{table*}

\begin{table*}
\centering
\caption{Predictions for the number of associated and unassociated sources in the {\bf unweighted}  training catalog.
For the definition of the classes, see Table \ref{tab:classes}.
}
\label{tab:summary_unweighted}
\begin{tabular}{lllllll}
\hline
 & Class label & N assoc & RF assoc & NN assoc & RF unas & NN unas \\
\hline
1 & spp+ & 173 & $175.5\pm 2.0$ & $170.7\pm 1.5$ & $420.4\pm 3.1$ & $448.9\pm 2.4$ \\
2 & msp+ & 179 & $177.5\pm 1.7$ & $181.3\pm 2.0$ & $182.3\pm 2.0$ & $197.0\pm 1.7$ \\
3 & psr+ & 241 & $236.2\pm 2.0$ & $238.8\pm 2.2$ & $187.6\pm 2.1$ & $181.7\pm 1.6$ \\
4 & fsrq+ & 827 & $829.8\pm 3.5$ & $828.3\pm 2.3$ & $200.5\pm 2.1$ & $185.7\pm 1.2$ \\
5 & bcu+ & 1637 & $1641.8\pm 4.9$ & $1619.0\pm 3.1$ & $1253.3\pm 4.1$ & $1269.7\pm 2.9$ \\
6 & bll+ & 1557 & $1553.2\pm 4.2$ & $1576.0\pm 3.0$ & $332.9\pm 2.5$ & $294.0\pm 1.6$ \\
\hline
\end{tabular}
\end{table*}

\begin{figure*}
\includegraphics[width=\textwidth]{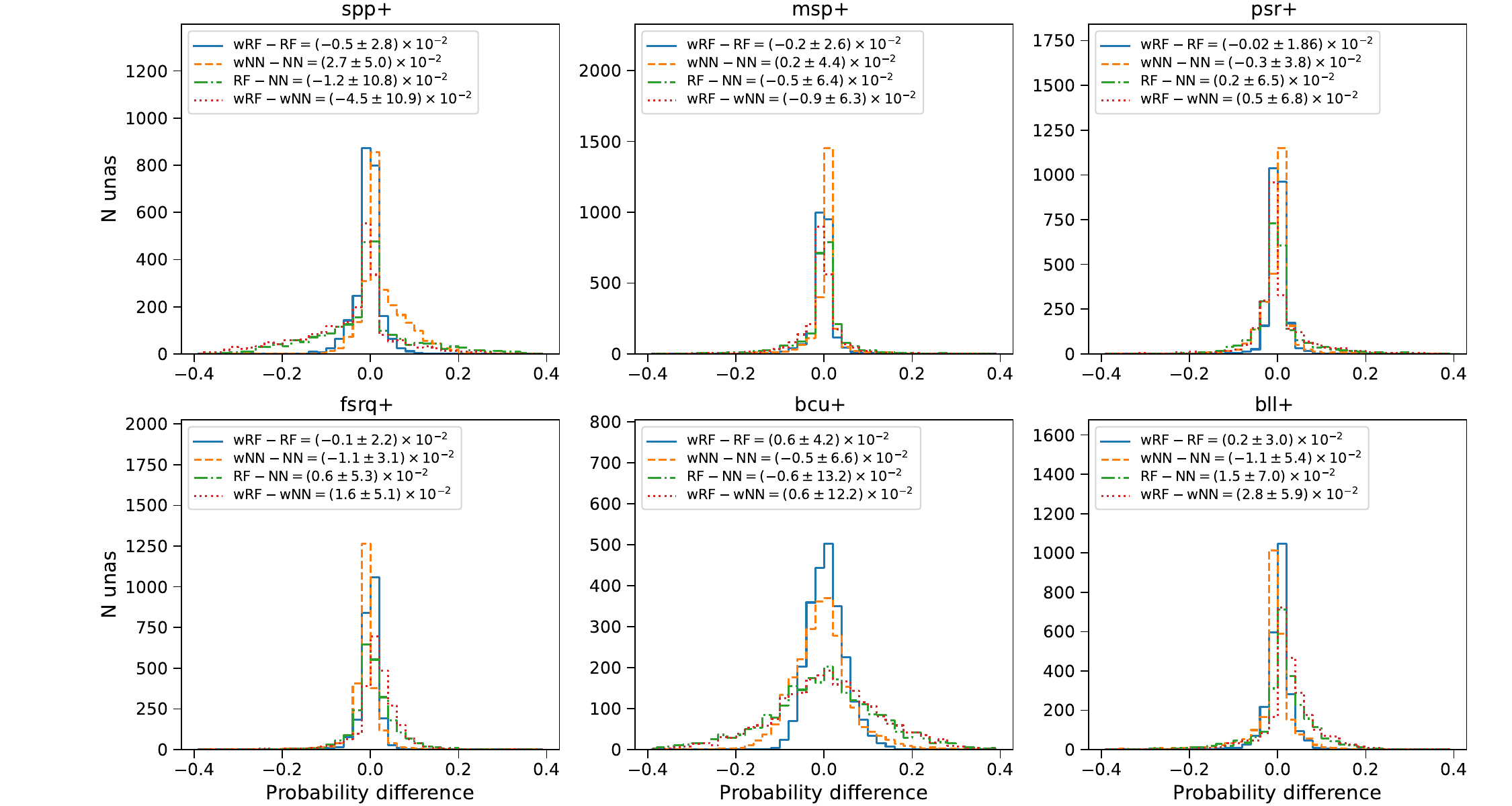}
\caption{
Difference of class probabilities for individual sources determined with RF (``RF'' labels) and NN (``NN'' labels) algorithms using weighted (``w'' is included in labels)
and unweighted training.
The corresponding overall mean difference and the standard deviations are reported in the labels.
}
\label{fig:prob_comparisons}
\end{figure*}

\begin{figure*}
\includegraphics[width=\textwidth]{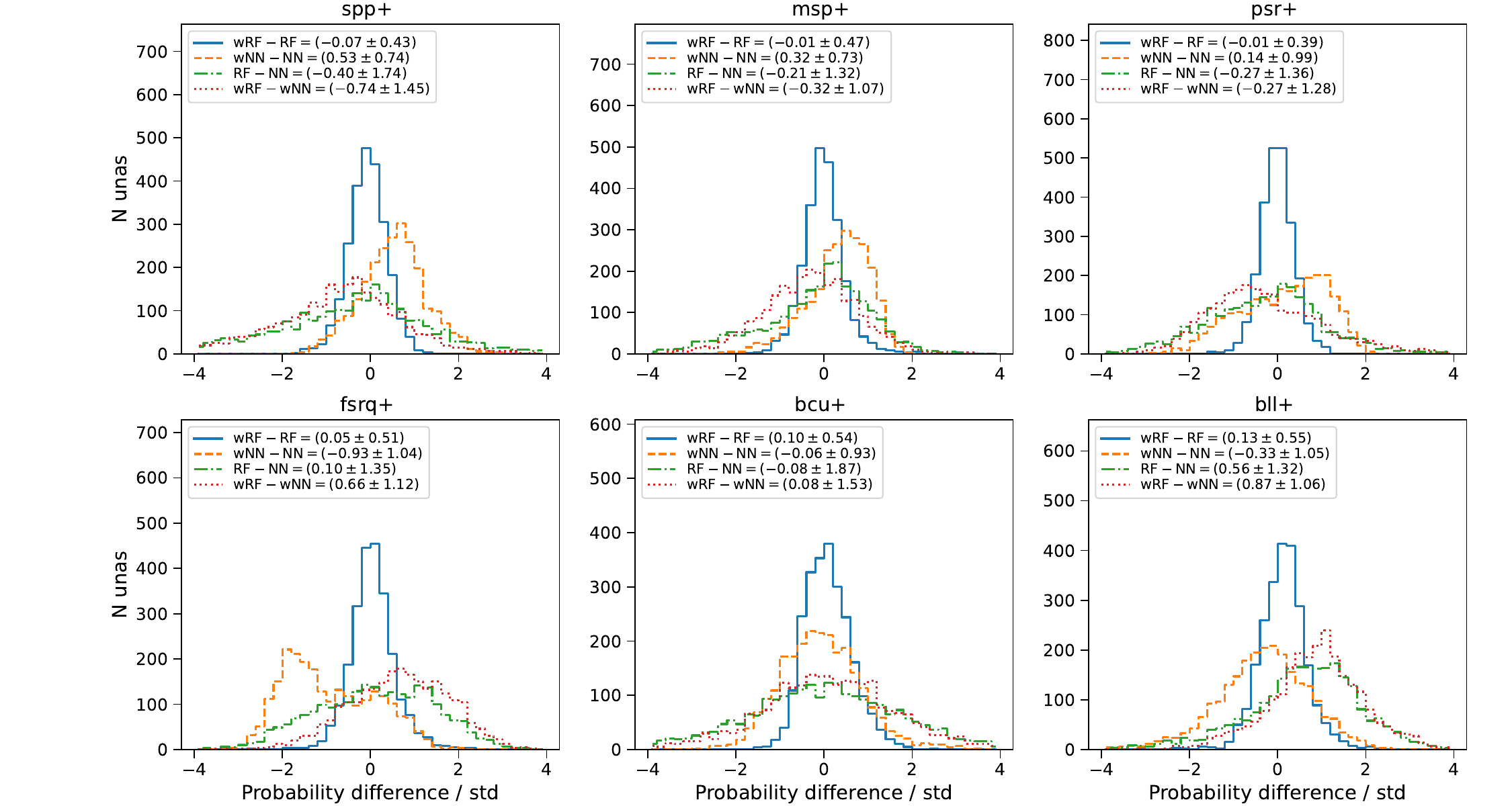}
\caption{
Difference of class probabilities for individual sources determined with RF (``RF'' labels) and NN (``NN'' labels) algorithms using weighted (``w'' is included in labels)
and unweighted training relative to the standard deviations of the predicted probabilities (see text for more details).
The corresponding overall mean relative difference and the standard deviations of the relative differences are reported in the labels.
}
\label{fig:prob_comparisons_rel}
\end{figure*}

We compare the changes in the individual class probabilities for the unassociated sources for weighted and unweighted training in 
Figs. \ref{fig:prob_comparisons} and \ref{fig:prob_comparisons_rel}.
In Fig.~\ref{fig:prob_comparisons} we calculate the difference of predicted class probabilities in four cases:
weighted RF minus unweighted RF probability (``wRF -- RF'' label),
weighted NN minus unweighted NN probability (``wNN -- NN'' label),
unweighted RF minus unweighted NN probability (``RF -- NN'' label),
and weighted RF minus weighted NN probability (``wRF -- wNN'' label).
We see that in all cases the differences of the class probabilities for the individual sources is within about 13\%.
The smallest differences (within about 4\%) are among weighted RF and unweighted RF probabilities for all classes.
The largest standard deviations of about 13\% are for the RF minus NN probabilities for the unweighted and weighted trainings
in the bcu+ class.

\begin{figure*}
\includegraphics[width=\cmsize\columnwidth]{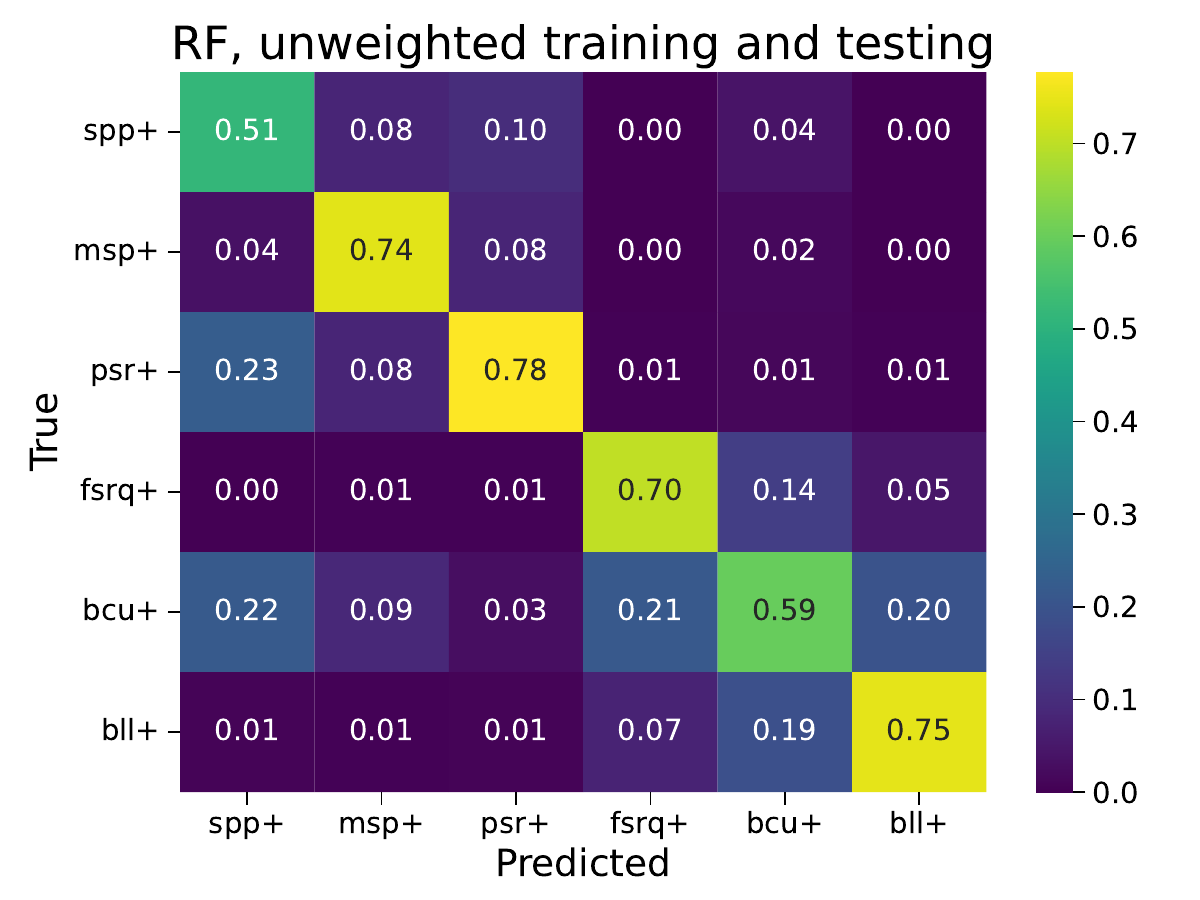}
\includegraphics[width=\cmsize\columnwidth]{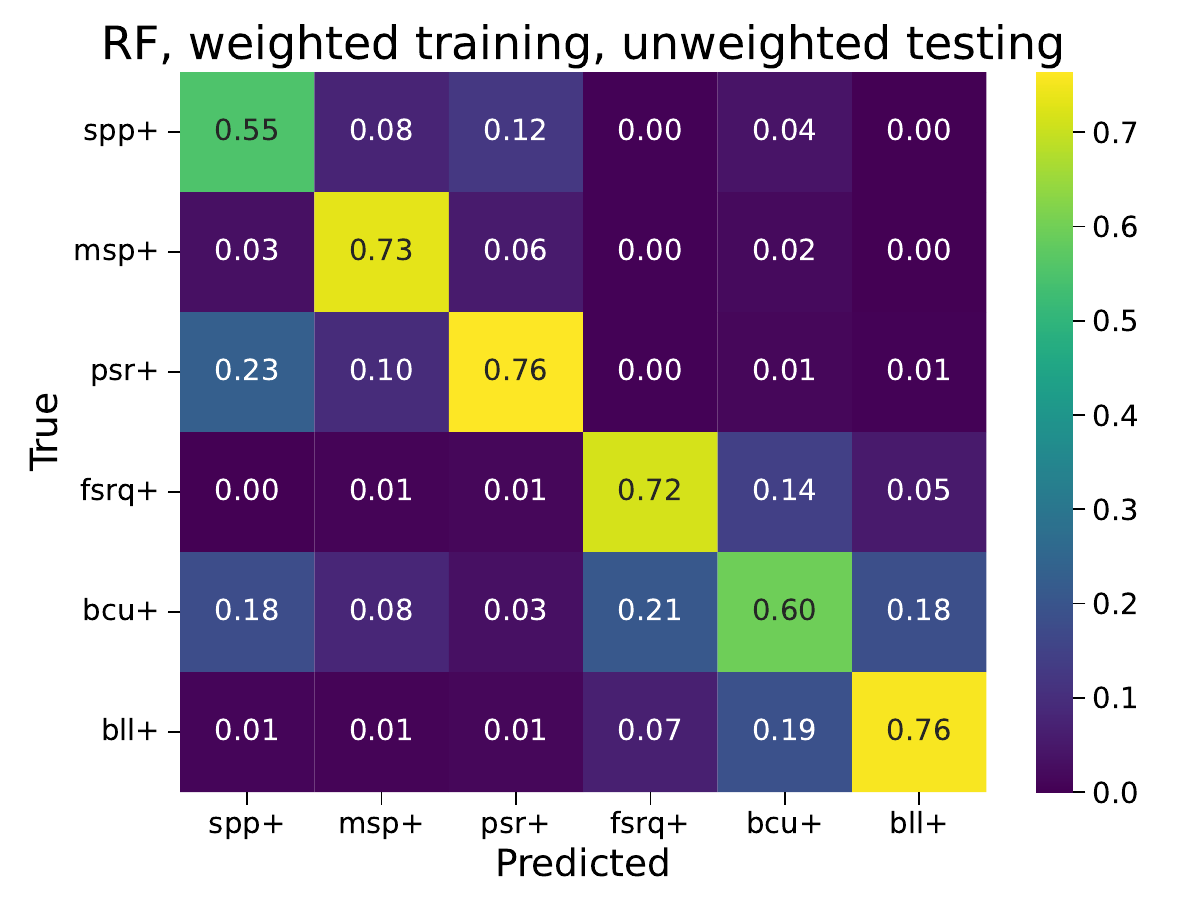} \\
\includegraphics[width=\cmsize\columnwidth]{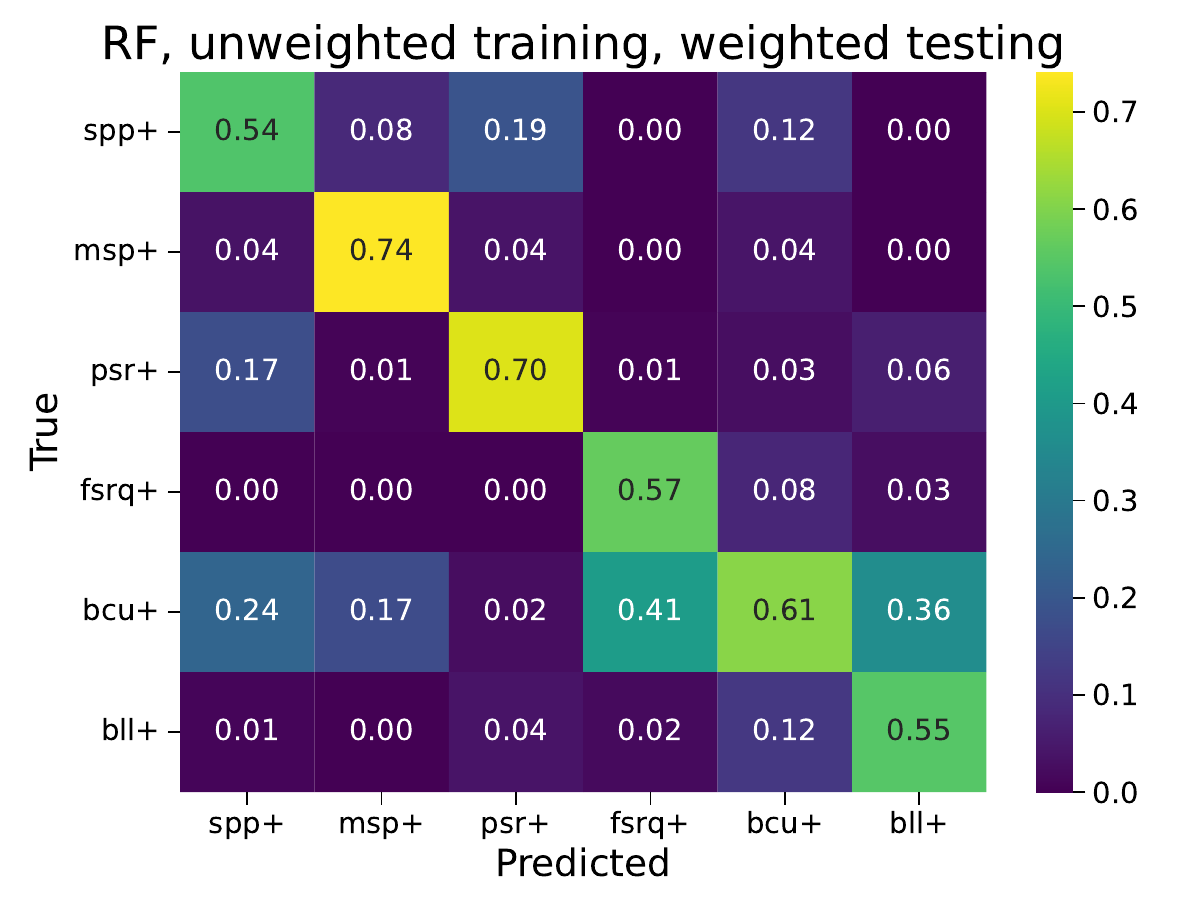}
\includegraphics[width=\cmsize\columnwidth]{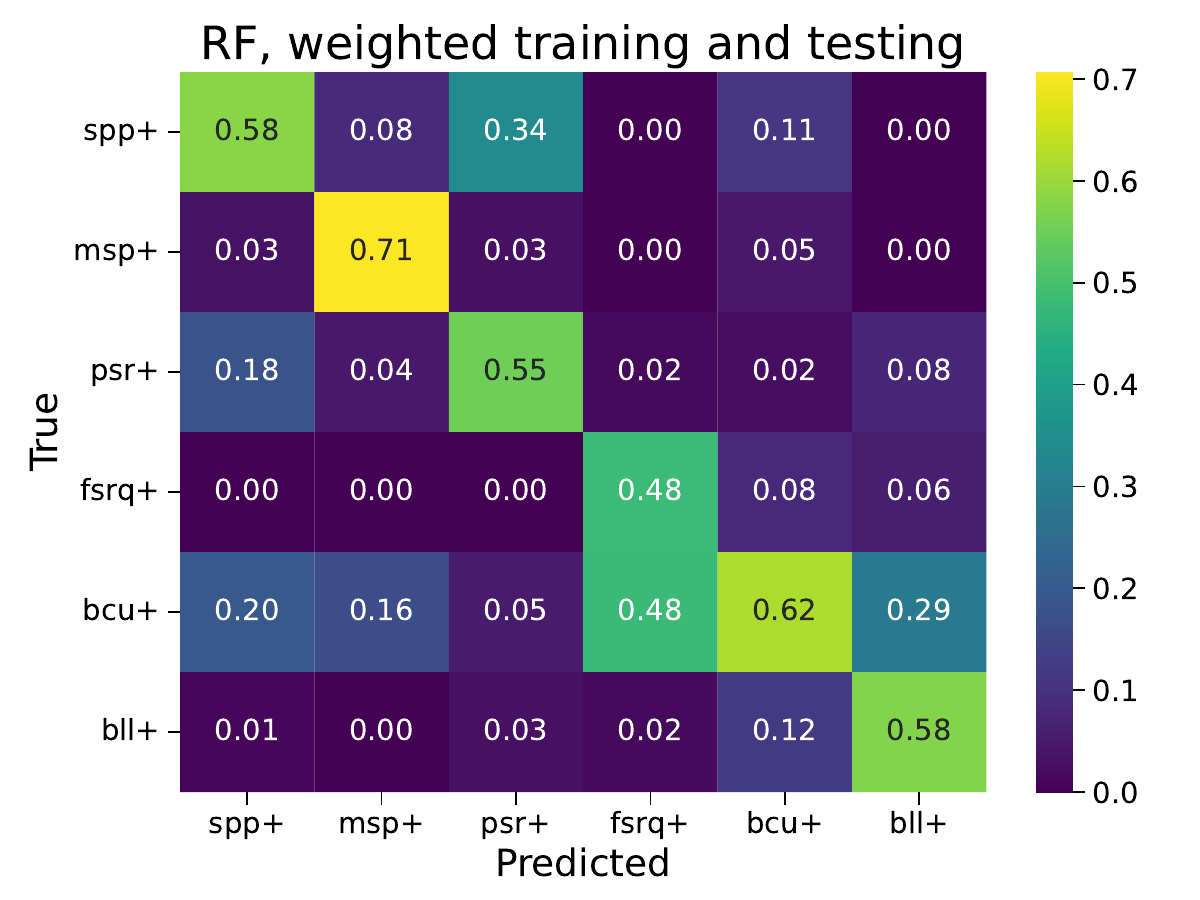}
\caption{
Confusion matrix normalized to the number of predicted sources in a class. The sources are classified according to the highest class probability for each source.
The numbers on the diagonal show the one-vs-all precision for this classification method. Top (bottom) panels show the performance for the unweighted (weighted) samples representative for the associated (unassociated) sources.
}
\label{fig:CM_RF}
\end{figure*}

Fig.~\ref{fig:prob_comparisons_rel} is similar to Fig.~\ref{fig:prob_comparisons} but we divide the difference of the probabilities
for each source by the RMS of the uncertainties due to random training/testing splits, i.e., we plot the histograms of
$
{(p_1 - p_2)}/{\sqrt{(\sigma^2_1 + \sigma^2_2) / 2}}.
$
We see that in most cases the difference in class probabilities for the individual sources for different classification methods is comparable to the standard deviations
due to training / testing splits (ranging from about 0.5 to 2 sigma).
We create probabilistic catalogs constructed both with unweighted and with weighted training samples.

In Fig. \ref{fig:CM_RF} we compare the confusion matrices for the RF classification with the weighted and unweighted training and testing datasets.
The predicted classes are calculated by taking the class with the largest probability for each source.
In this calculation we take all associated sources and use the class probabilities determined as a mean over the training / testing splits when the sources are in the testing datasets.
We see that testing with unweighted samples gives similar performance estimates both for 
training on unweighted samples (top left panel) and for training on weighted samples (top right panel).
While for testing with weighted samples, training with 
unweighted samples (bottom left panel) has a better performance for psr+ and fsrq+ classes compared to the 
training with weighted samples (bottom right panel).

\section{Conclusions}
\label{sec:conclusions}

In the paper we study the effect of the covariate shift (due to difference in the distributions of 
associated and unassociated sources) on the multi-class classification of \Fermi-LAT sources.
In order to realistically estimate the expected performance for the classification of unassociated sources using only the associated sources,
we introduce sampling weights proportional to the ratio of the PDFs for the unassociated sources to the associated sources,
so that the PDF of associated sources weighted by these sampling weights is similar to the PDF of the unassociated ones.

We use RF and NN algorithms and perform training with both weighted and unweighted samples.
We test the performance using weighted testing samples drawn from the associated sources, which is expected
to give realistic estimates of the performance for the unassociated sources.
We find that
\ben
\item
{\bf Covariate shift has little effect on estimated class probabilities for individual sources.}
The difference among class probabilities for individual unassociated sources derived with RF and NN algorithms using either weighted or unweighted training 
samples are comparable to the statistical uncertainties for the probabilities estimated from the random splits into training and testing datasets.
This result justifies the use of unweighted training samples in the derivation of the classification algorithms,
which are then used to classify unassociated sources.
\item
{\bf Using weighted or unweighted training samples has little effect on average performance.}
The average performance, estimated using ROC curves, precision, recall, and reliability diagrams,
is similar for weighted and for unweighted training. 
However, the variance is observed to increase for NN algorithms in the weighted training case 
for some of the characteristics (e.g., precision and reliability) for classes with a significant decrease due to weights
in the effective sample size (e.g., FSRQs).
\item
{\bf Covariate shift results in a decrease of up to 20\% in precision and recall for some classes, 
estimated with weighted testing samples
compared to estimates with unweighted testing samples.}
The most affected classes are the classes of extra-galactic sources (such as FSRQs and BL Lacs) dominated by sources at high latitudes 
where the fraction of unassociated sources is smaller than at low latitudes.
\een

The overall conclusion is that both unweighted and weighted training have similar expected performance,
but the covariate shift should be taken into account in the estimations of the performance, e.g., with the weighted testing samples.
Weighted training can lead to larger variance of the expected performance (especially for classes with a significant reduction
in effective sample number) and it also leads to biased predictions for the unweighted tests.
As a result, we find that it is better to perform the classification with unweighted training dataset but, for a realistic estimate of the classification performance, one should use weighted testing samples.

We create probabilistic catalogs using RF and NN algorithms trained with weighted and unweighted samples.
The catalogs include predicted class probabilities for 6 classes with RF and NN algorithms averaged over 51 realizations
of the 70/30\% training/testing datasets (for associated sources the probabilities are averaged over the splits when the source appears in the testing sample)
as well as the standard deviations of the probabilities due to the splits.
We also add a column with the sample weights.
We calculate the expected number of sources in the six classes among the unassociated sources.
The largest fractional increase is expected for the spp+ class: there are about 2.5 times more expected 
spp+ sources among the unassociated ones than there are associated spp+ sources. 
For comparison, the expected number of msp+ (psr+) sources among the unassociated sources is about the number of (75\% of) associated msp+ (psr+) sources.
For the extra-galactic sources, the largest fractional increase is for the bcu+ sources: the expected increase is more than 70\%, compared to an increase 
of about 20\% for bll+ and fsrq+ classes.
The catalogs are publicly available at \zenodo.

\section*{Acknowledgements}

The author would like to thank Aakash Bhat, Bryan Zaldivar, and anonymous referees for important comments and suggestions, as well as acknowledge support by the DFG grant MA 8279/3-1
and the use of the following software:
Astropy \citep[\url{http://www.astropy.org},][]{2013A&A...558A..33A}, 
Matplotlib \citep[\url{https://matplotlib.org/},][]{Hunter:2007}, 
pandas \citep[\url{https://pandas.pydata.org/},][]{mckinney-proc-scipy-2010},
scikit-learn \citep[\url{https://scikit-learn.org/stable/},][]{scikit-learn},
and TensorFlow \citep[\url{https://www.tensorflow.org/},][]{tensorflow2015-whitepaper}.


\section*{Data Availability}

The results of this work are based on the publicly available \Fermi-LAT 4FGL-DR4 catalog \url{https://fermi.gsfc.nasa.gov/ssc/data/access/lat/14yr_catalog/}
\citep{2023arXiv230712546B}.
The results of this work are available online at \zenodo.
 



\bibliographystyle{rasti}
\bibliography{LAT_cov_shift_bibl} 



\appendix

\section{Model the distributions of associated and unassociated sources}
\label{app:GMM_Model}

\begin{figure}
\includegraphics[width=\columnwidth]{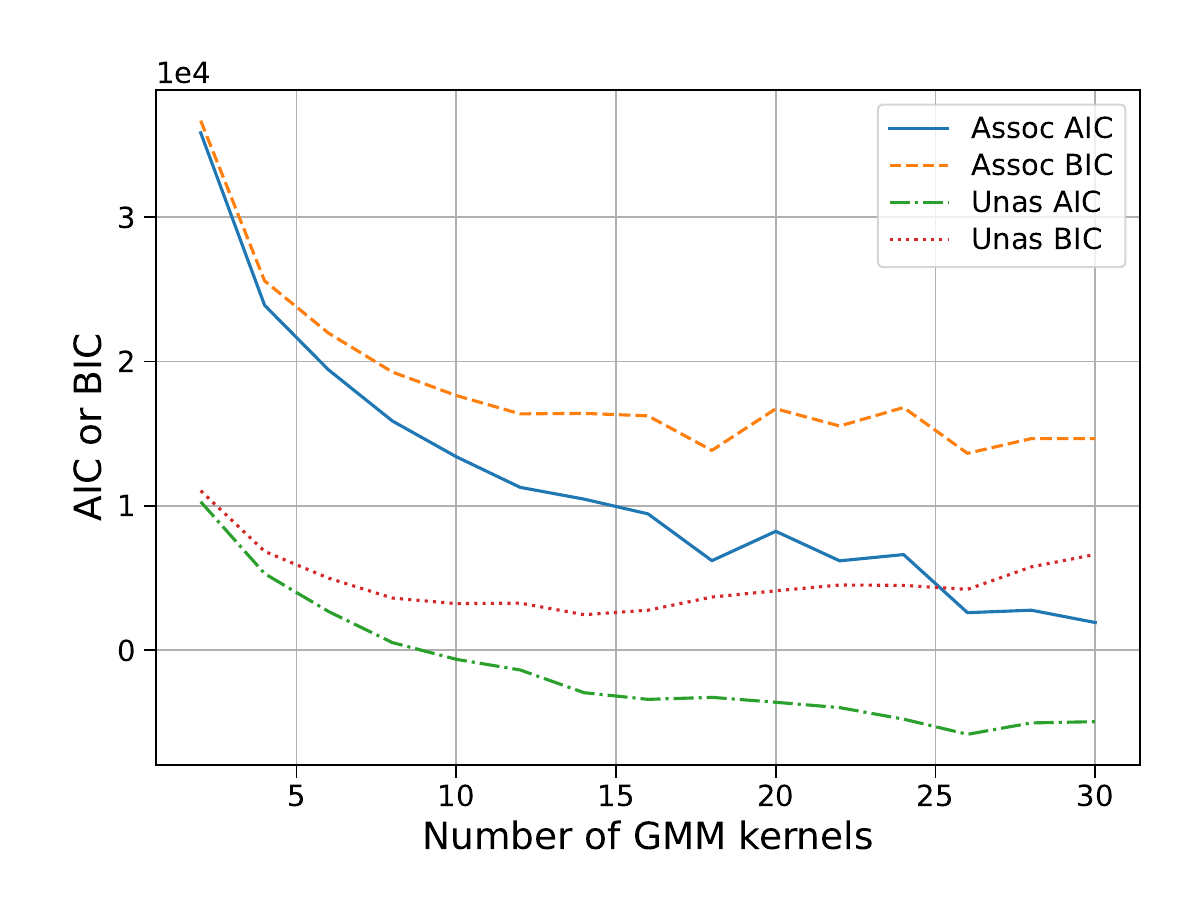}
\caption{
The Akaike \citep{1100705_AIC} and Bayesian \citep{10.1214/aos/1176344136_BIC}
information criteria as functions of the number of GMM kernels
for associated (unassociated) sources: ``Assoc AIC'' and ``Assoc BIC'' (``Unas AIC'' and ``Unas BIC'') labels respectively.
Both AIC and BIC are decreasing up to about 12 GMM kernels.
}
\label{fig:AIC_BIC}
\end{figure}

\begin{figure}
\includegraphics[width=\columnwidth]{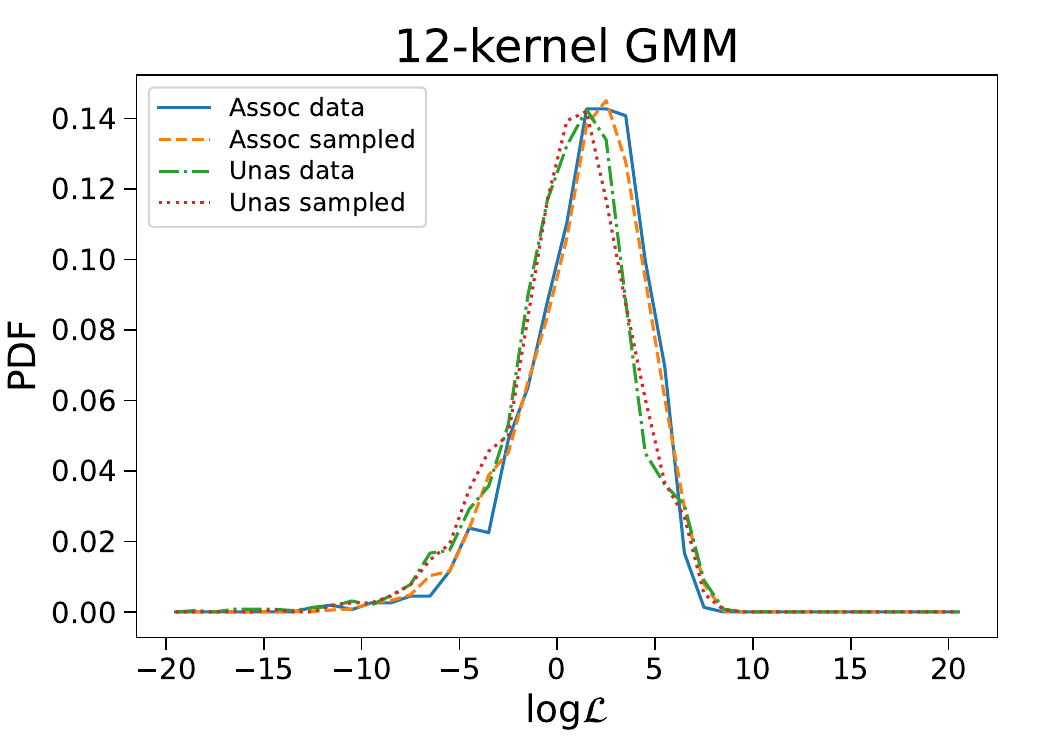}
\caption{
The distribution of log-likelihoods in the 12-kernel GMM models of the associated and unassociated sources.
Solid (dash-dotted) line: the distribution of log-likelihoods for associated (unassociated) sources.
Dashed (dotted) line: the distribution of log-likelihoods for sources sampled from the GMM model for 
associated (unassociated) sources.
}
\label{fig:logL_PDFs}
\end{figure}

In this appendix we construct PDFs of associated and unassociated sources using GMMs.
In order to determine an optimal number of kernels in the GMMs
we compare the Akaike information criterion \citep[AIC,][]{1100705_AIC}
\be
{\rm AIC} = 2 k  - 2 \ln \cal{L},
\ee
where $k$ is the number of parameters in a model and $\cal{L}$ is the likelihood of the model,
and the Bayesian information criterion \citep[BIC,][]{10.1214/aos/1176344136_BIC}
\be
{\rm BIC} = k \ln n - 2 \ln \cal{L},
\ee
where $k$ and $\cal{L}$ are the same as in the AIC and $n$ is the number of data samples.
In Fig.~\ref{fig:AIC_BIC} we plo thet AIC and BIC as functions of the number of the GMM kernels.
We use the AIC and BIC implementations in the scikit-learn package \citep{scikit-learn}.
We see that above about 12 GMM kernels BIC is approximately constant for both associated and unassociated sources.
Although AIC continues to decrease above 12 kernels, we choose the GMM models with 12 kernels as the models with minimal complexity
up to which both AIC and BIC are decreasing.
We also test in Fig.~\ref{fig:logL_PDFs} that for the 12-kernel GMM models
the distributions of likelihoods for the associated and unassociated sources are
similar to the distributions of likelihoods for the samples drawn from models for the associated and unassociated sources respectively.

\section{Selection of input features and feature importance}
\label{app:features}

\begin{table}
\centering
\caption{Feature importance for RF classification with unweighted training and testing for
different numbers of classes corresponding to the different depth of the class separation in the bottom 
panel of Fig. \ref{fig:class_def}. Depths 1, 2, 3, and 4 correspond to 2, 4, 5 and 6-class classifications respectively.}
\label{tab:RF_features}
\begin{tabular}{lrrrr}
\hline
Feature & Depth 1 & Depth 2 & Depth 3 & Depth 4 \\
\hline
LP\_index1000MeV & 0.159 & 0.167 & 0.160 & 0.163 \\
log10(Signif\_Avg) & 0.077 & 0.128 & 0.140 & 0.142 \\
log10(Variability\_Index) & 0.100 & 0.114 & 0.114 & 0.113 \\
log10(Unc\_Energy\_Flux100) & 0.128 & 0.109 & 0.105 & 0.108 \\
LP\_beta & 0.103 & 0.106 & 0.105 & 0.101 \\
log10(Energy\_Flux100) & 0.112 & 0.089 & 0.092 & 0.092 \\
sin(GLAT) & 0.055 & 0.081 & 0.087 & 0.087 \\
LP\_SigCurv & 0.164 & 0.097 & 0.082 & 0.081 \\
cos(GLON) & 0.052 & 0.056 & 0.058 & 0.059 \\
sin(GLON) & 0.051 & 0.054 & 0.057 & 0.055 \\
\hline
\end{tabular}
\end{table}

In this work we use 10 input features (described in Section \ref{sec:data}).
Although \Fermi-LAT catalogs have many more source parameters,
most of these parameters are highly correlated \citep{2022A&A...660A..87B}.
In particular, there are different representations of the same quantity, such as the position on the sky
in Galactic or equatorial coordinates. There are also high correlations, e.g., among uncertainties in flux, energy flux,
and spectrum normalization \citep{2022A&A...660A..87B}.
It has been noted by \cite{2020MNRAS.492.5377L} that, in case of two-class classification,
using more than five input features does not significantly 
improve the classification performance with the increasing complexity of the model.

In this work we use the features previously selected by \cite{2020MNRAS.492.5377L} and \cite{2022A&A...660A..87B}
with several modifications:
(i) instead of the hardness ratios, we use the log-parabola curvature parameter to describe the change of the ``spectral index'' as a function of energy;
(ii) instead of the spectral index parameter (or power-law spectral index), we use the index of the log-parabola spectrum at 
1 GeV, which is independent of the pivot energy and is well defined for curved spectra;
(iii) we add the average significance parameter (``Signif\_Avg''). Although this parameter is highly correlated with the energy flux above 100 MeV \citep{2022A&A...660A..87B}, ``log10(Signif\_Avg)'' has a higher importance than ``log10(Energy\_Flux100)'' in RF classifications with more than two classes (cf. Table \ref{tab:RF_features});
(iiii) we replace Galactic longitude with two parameters ``cos(GLON)'' and ``sin(GLON)'' in order to avoid discontinuity between $0^\circ$ and $360^\circ$.
The same input features have been previously used by \cite{2023MNRAS.521.6195M} in the determination of the hierarchical classification of the \Fermi-LAT sources.

We show the importance of the 10 input features in the RF classification (in the unweighted training case) in Table \ref{tab:RF_features}.
The features are ordered according to the importance for the six-class classification (the ``Depth 4'' column).
It is interesting to note that some features
have a significantly different importance in the two-class and in the multi-class classifications.
For instance, the spectral curvature significance parameter (``LP\_SigCurv'') has the highest significance for the two-class classification 
\citep[it has also been the most significant parameter in the analysis of][]
{2016ApJ...820....8S, 2020MNRAS.492.5377L, 2022A&A...660A..87B},
but it is in the middle (near the end) of the list for the four-class (five- and six-class) classification.
The parameter ``LP\_index1000MeV'' is the most important parameter for all classifications in this work,
while ``Spectral\_Index'' has been less significant in the previous two-class classifications, e.g.,
on place three in \cite{2020MNRAS.492.5377L} or on place four in \cite{2016ApJ...820....8S}, and log-parabola index
``LP\_Index'' was near the end of the significance table in \cite{2023MNRAS.521.6195M}.
On the other hand, parameters ``log10(Variability\_Index)'' and ``log10(Unc\_Energy\_Flux100)''
have similar importance for all cases (places three and four respectively) in this work as well as in the previous
analyses, where the importance of these features has been between place two and four \citep{2016ApJ...820....8S, 2020MNRAS.492.5377L, 2022A&A...660A..87B}.

\section{Neural networks}
\label{app:NNs}

\begin{figure}
\includegraphics[width=\columnwidth]{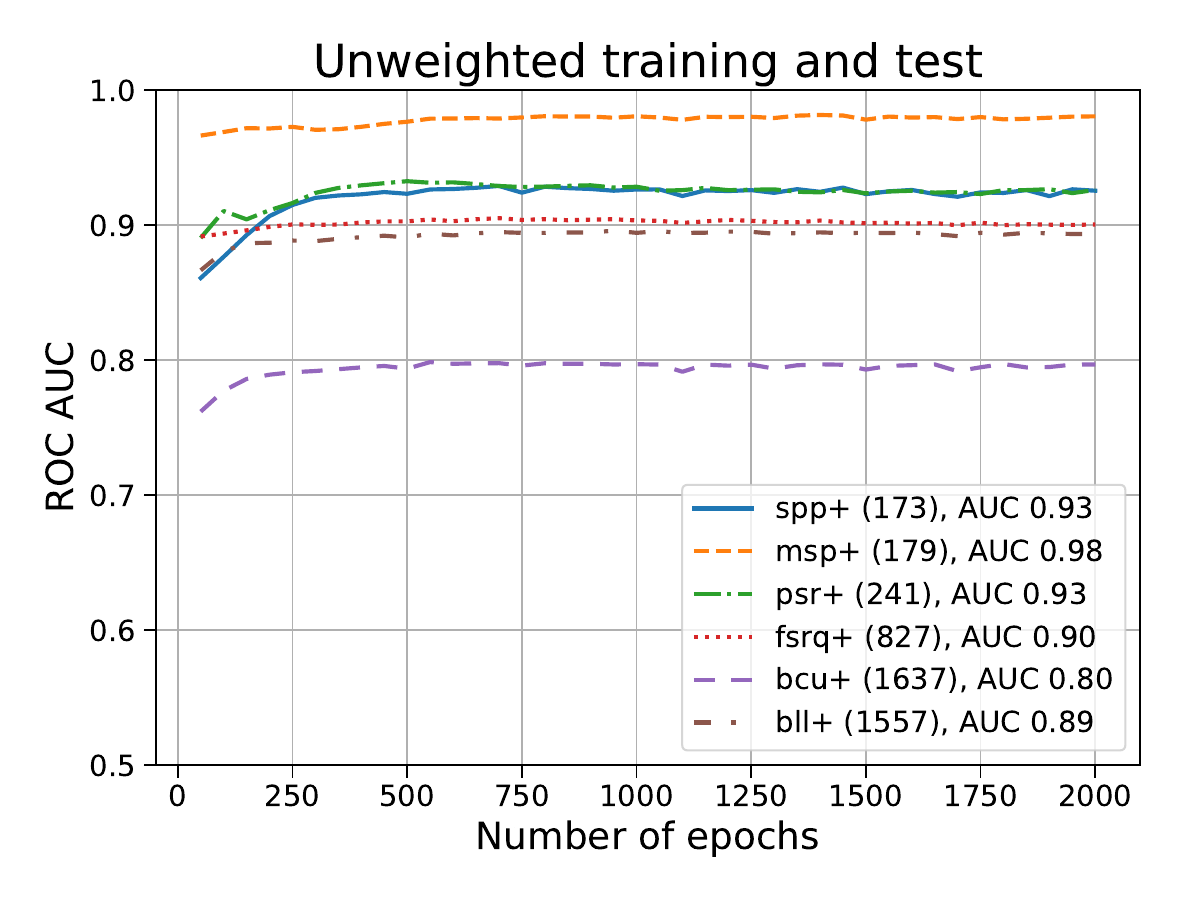} \\ 
\includegraphics[width=\columnwidth]{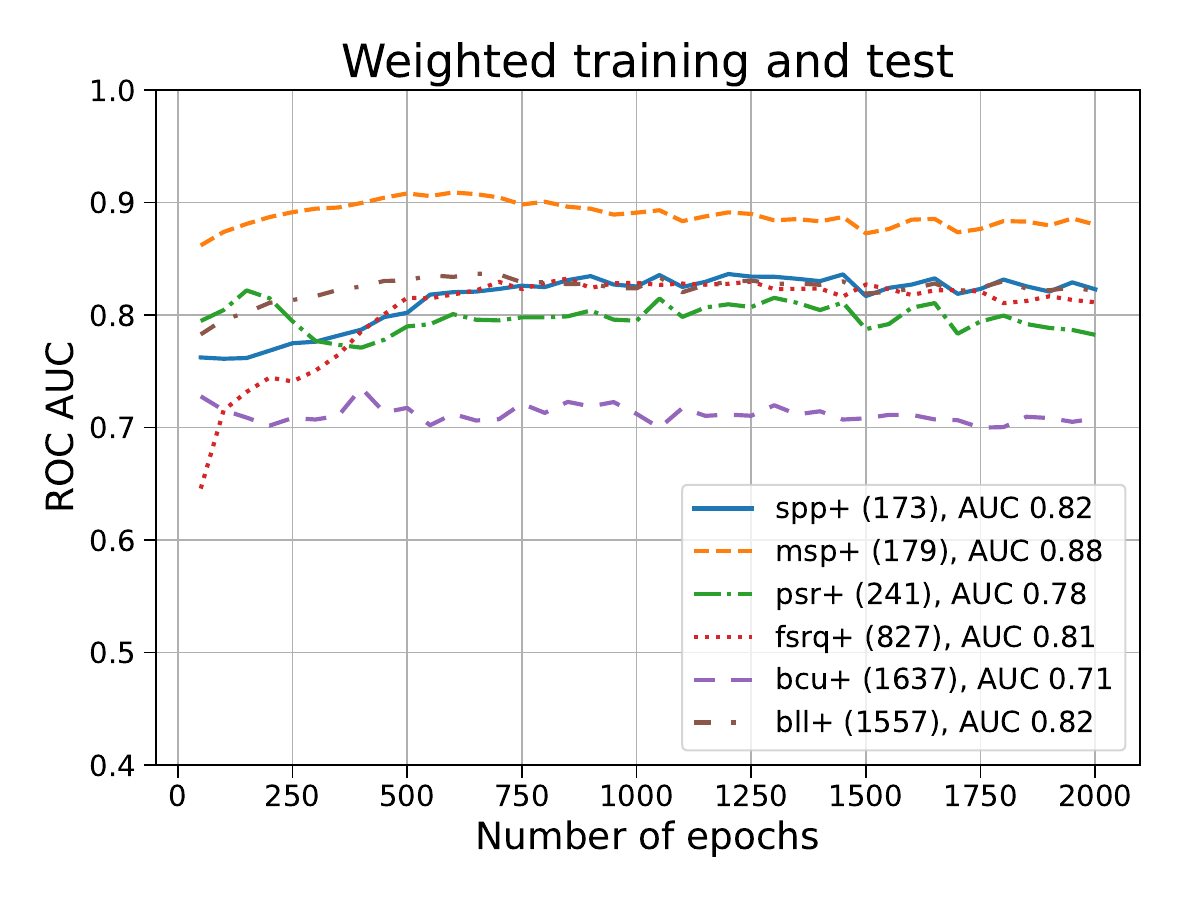}
\caption{
The ROC AUC for one-vs-all classification as a function of the number of epochs for the NN algorithm with unweighted (top panel) and weighted (bottom panel) 
samples.
}
\label{fig:NN_TF_AUC_test}
\end{figure}

In this appendix we provide details about the neural network method for the classification of \Fermi-LAT sources.
We use the TensorFlow implementation of neural networks \citep{tensorflow2015-whitepaper}.
We use stochastic gradient descent (adam) with learning rate of 0.001, two hidden layers with 20 and 10 nodes respectively, 
tanh activation functions, batch size of 200,
L2 regularization with $l2 = 0.001$, and no drop out.
We use the sparse categorical cross entropy loss function.
In Fig.~\ref{fig:NN_TF_AUC_test} we show the one-vs-all ROC AUC values for the six groups for the unweighted (top panel) and weighted (bottom panel) training.
We find that for the unweighted training there is no sign of overfitting up to the maximal number of epochs used in this test, e.g., 2000.
Provided that there is still a slight increase in performance in some groups up to about 500 epochs, we use 500 epochs for training the NN algorithm in the unweighted case.
For the weighted case, there are signs of overfitting for some groups above 500 epochs, e.g., for psr+ and fsrq+ groups.
As a result, we also use 500 epochs for the training in the weighted samples case.

\begin{figure*}
\includegraphics[width=\threesize\textwidth]{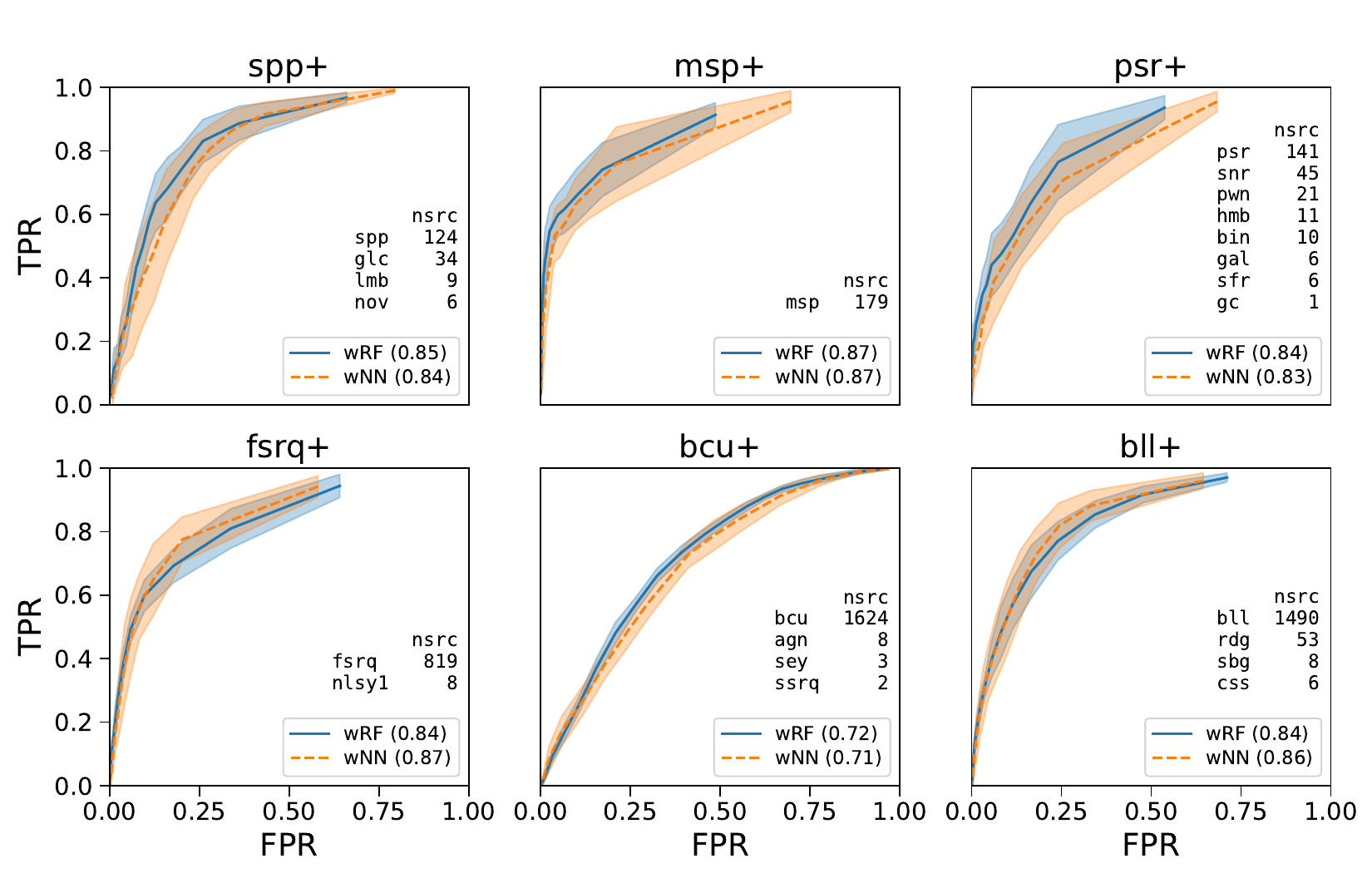}
\includegraphics[width=\threesize\textwidth]{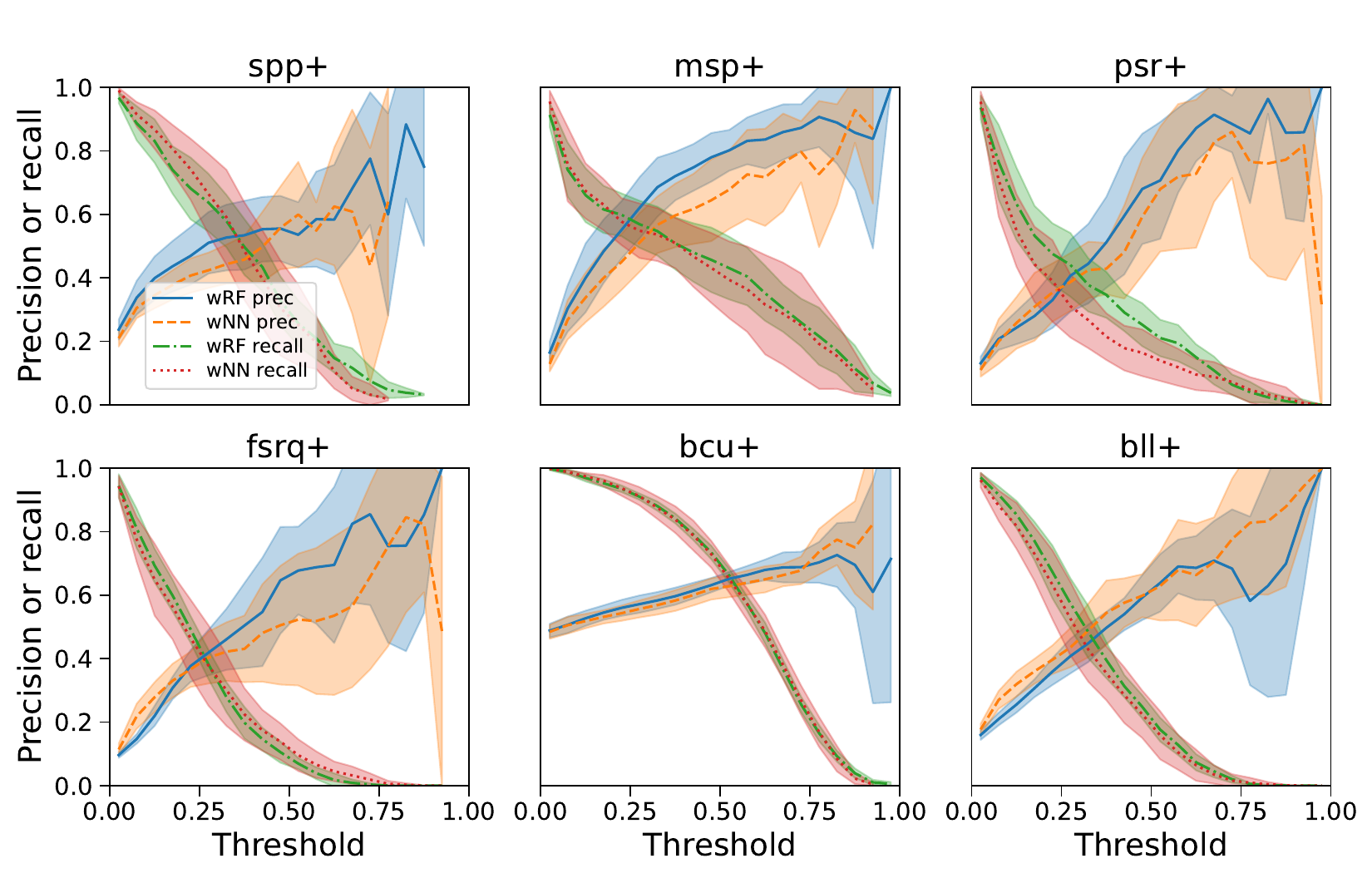}
\includegraphics[width=\threesize\textwidth]{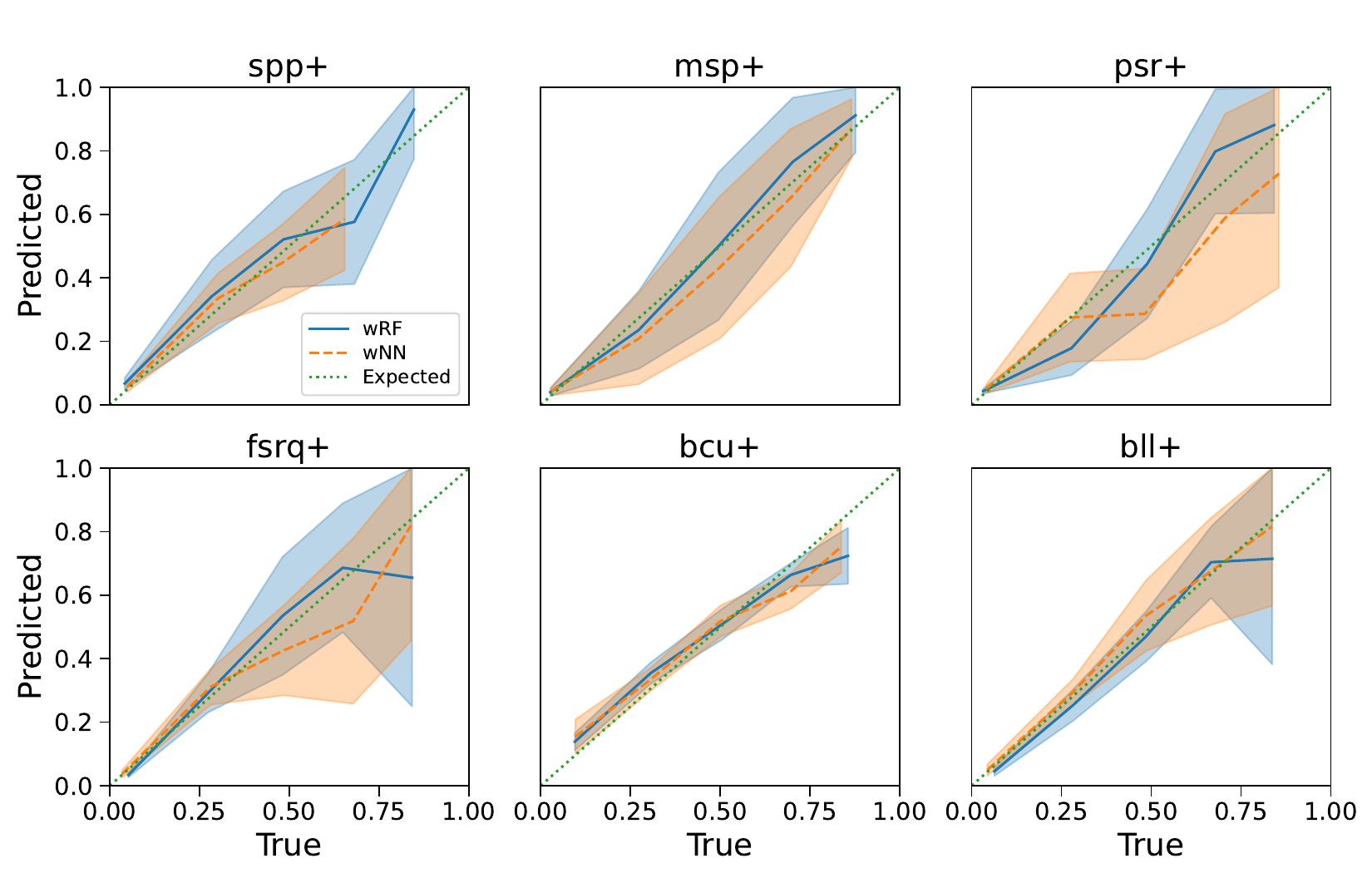}
\caption{
Comparison of ROC curves (top panels), precision and recall (middle panels), and calibration diagrams (bottom panels)
for training and testing with weighted samples using RF (``wRF'' labels) and NN (``wNN'' labels) algorithms.
The numbers in parenthesis in the top panel show the ROC AUC.
}
\label{fig:wRF_wNN_TF}
\end{figure*}

In Fig. \ref{fig:wRF_wNN_TF} we compare ROC curves, precision, recall, and calibration diagrams for
the training with weighted samples using NN and RF algorithms.
Generally, the performance of the NN is comparable to the performance of the RF.
RF gives slightly better results for the psr+ class, while NN has a better performance for the bll+ class.


\begin{figure*}
\includegraphics[width=\threesize\textwidth]{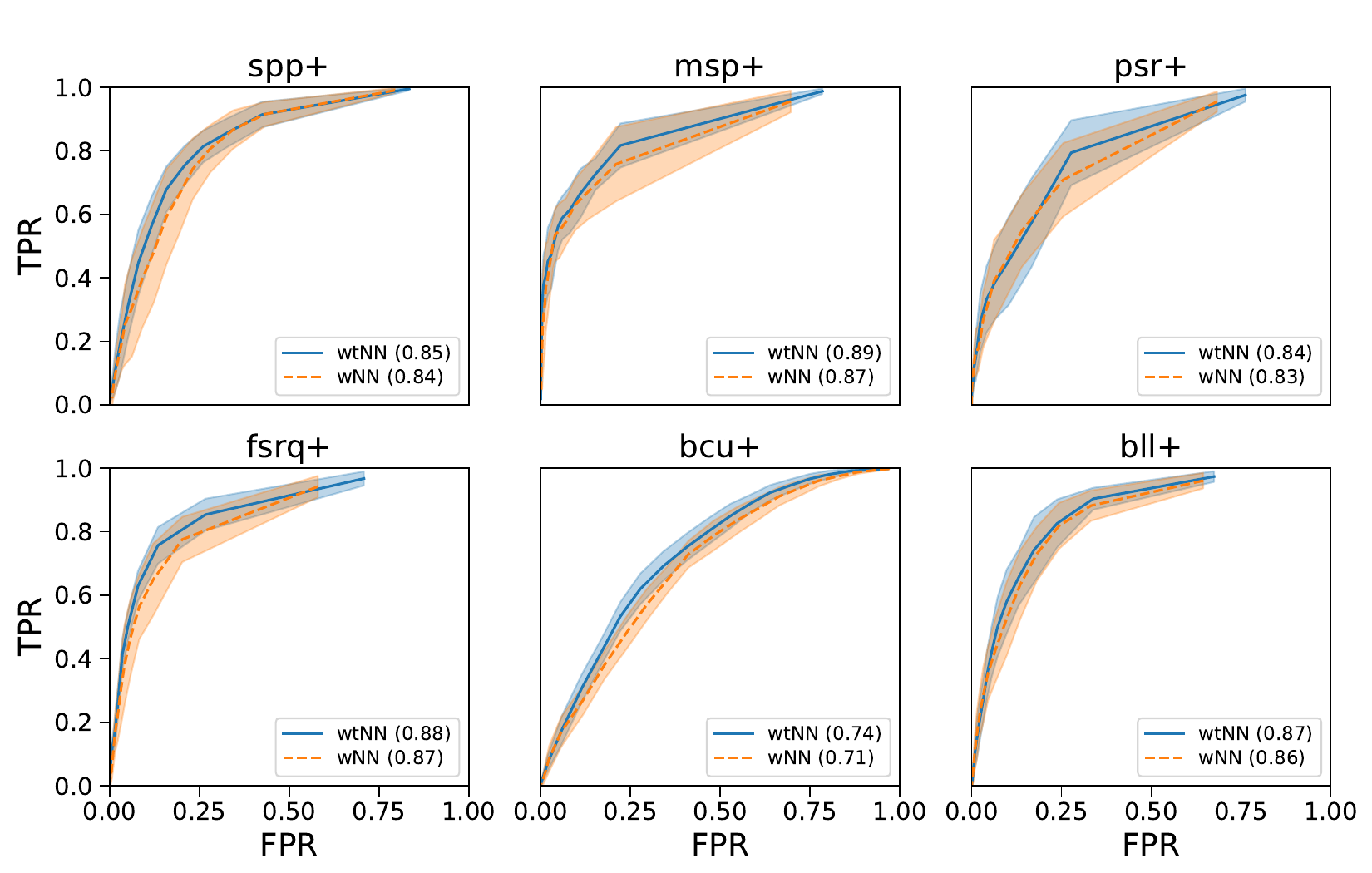}
\includegraphics[width=\threesize\textwidth]{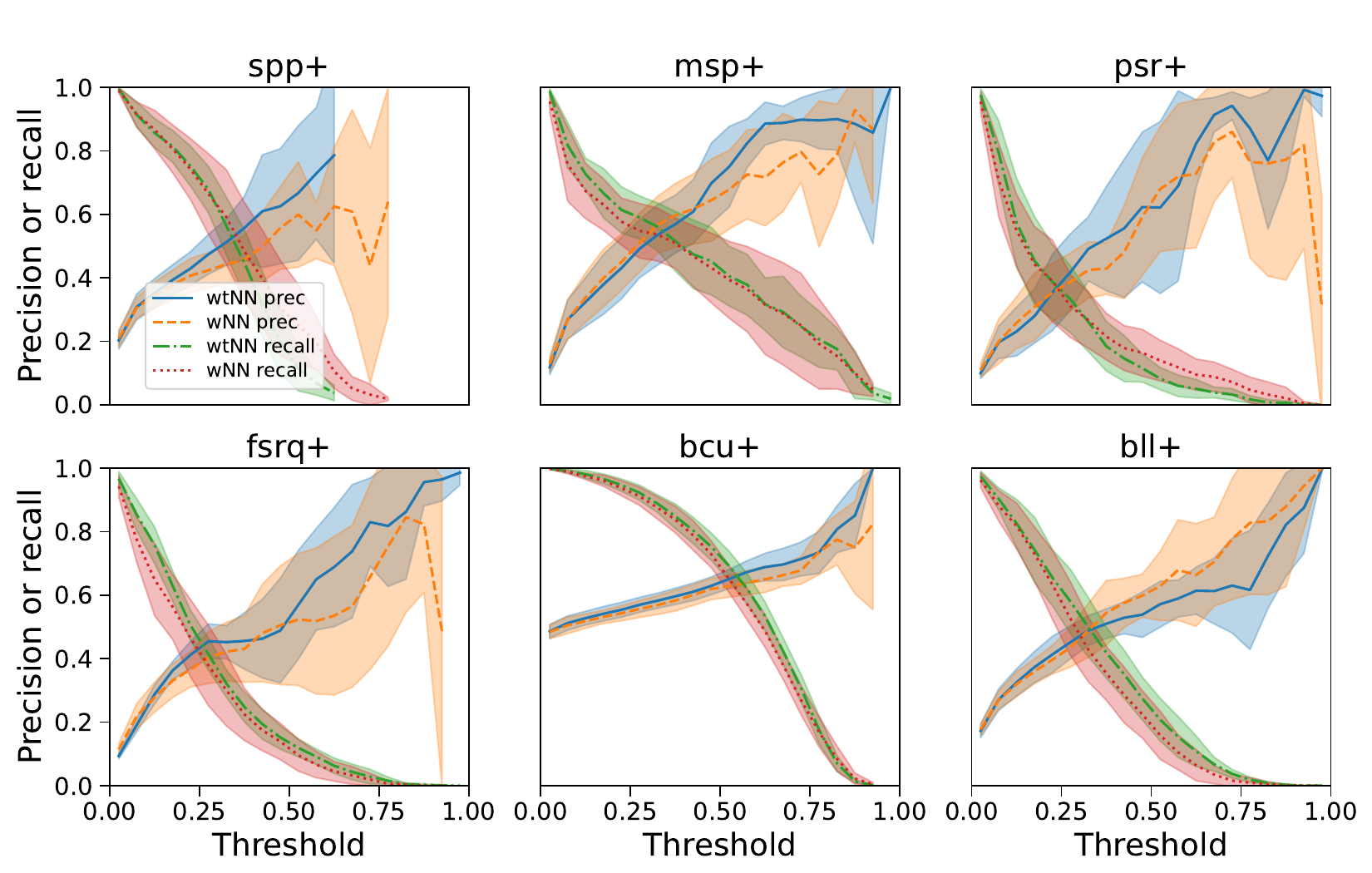}
\includegraphics[width=\threesize\textwidth]{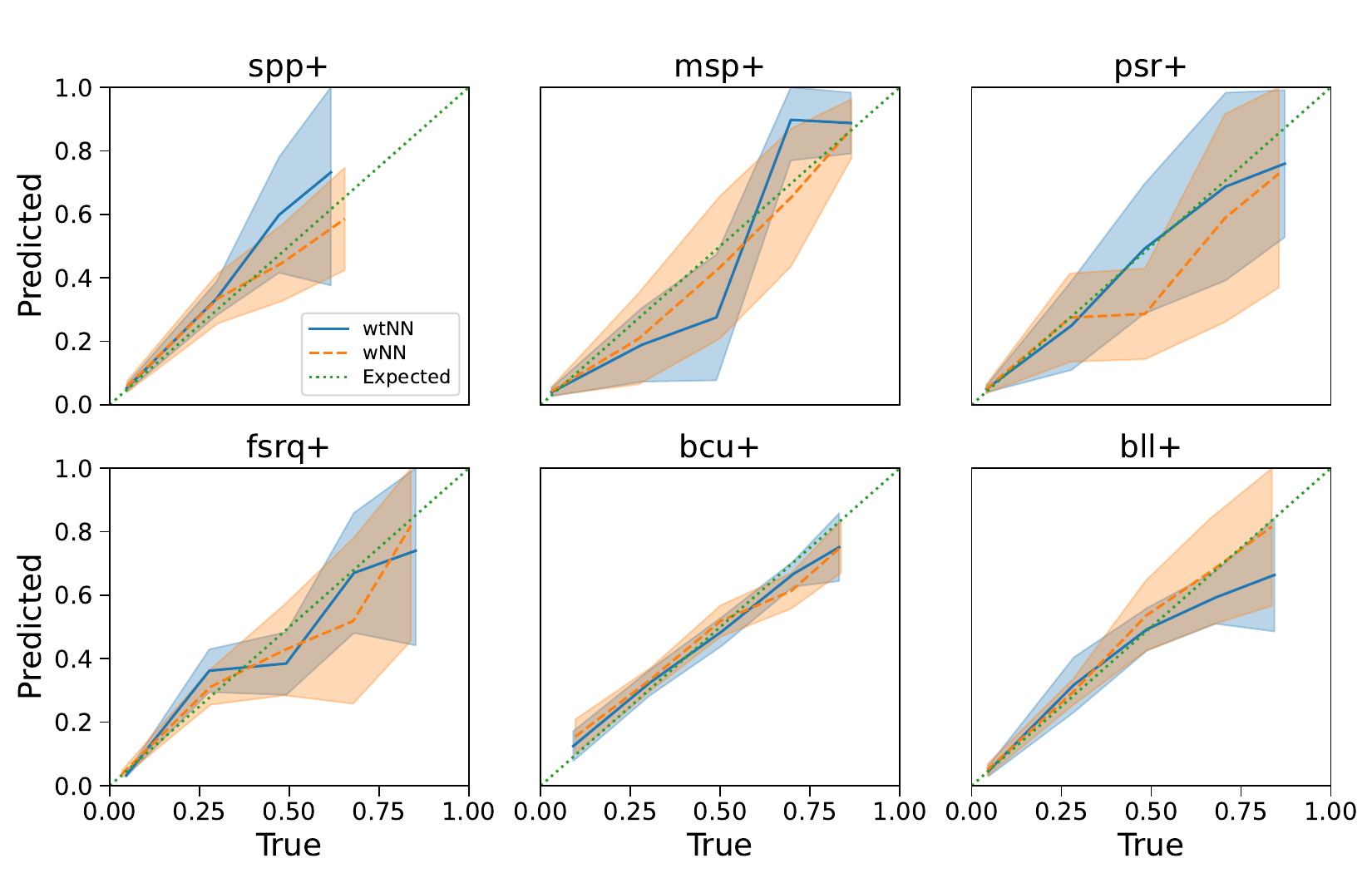}
\caption{
Comparison of ROC curves (top panels), precision and recall (middle panels), and calibration diagrams (bottom panels)
for classification with NN algorithms in case of 
training with unweighted samples and testing with weighted samples (``wtNN'' labels) and with weighted samples used both for
training and testing (``wNN'' labels).
}
\label{fig:wtNN_TF_wNN_TF}
\end{figure*}

In Fig. \ref{fig:wtNN_TF_wNN_TF} we compare the performance of the NN algorithm 
trained with unweighted and tested with weighted samples (``wtNN'' labels)
vs trained and tested with weighted samples (``wNN'' labels).
Most of the characteristics are similar for training with weighted and unweighted samples.
However, the statistical uncertainty band is narrower in the unweighted training case
for the fsrq+ and bll+ classes for precision (middle panels) and reliability (bottom panels).
This can be attributed to the fact that in the weighted samples case the ratio of the effective number of samples 
to the number of associated sources in
the fsrq+ and bll+ classes is the smallest among the six classes, which leads to an increased variance for this class in the weighted training case compared to the unweighted training.

\begin{figure*}
\includegraphics[width=\cmsize\columnwidth]{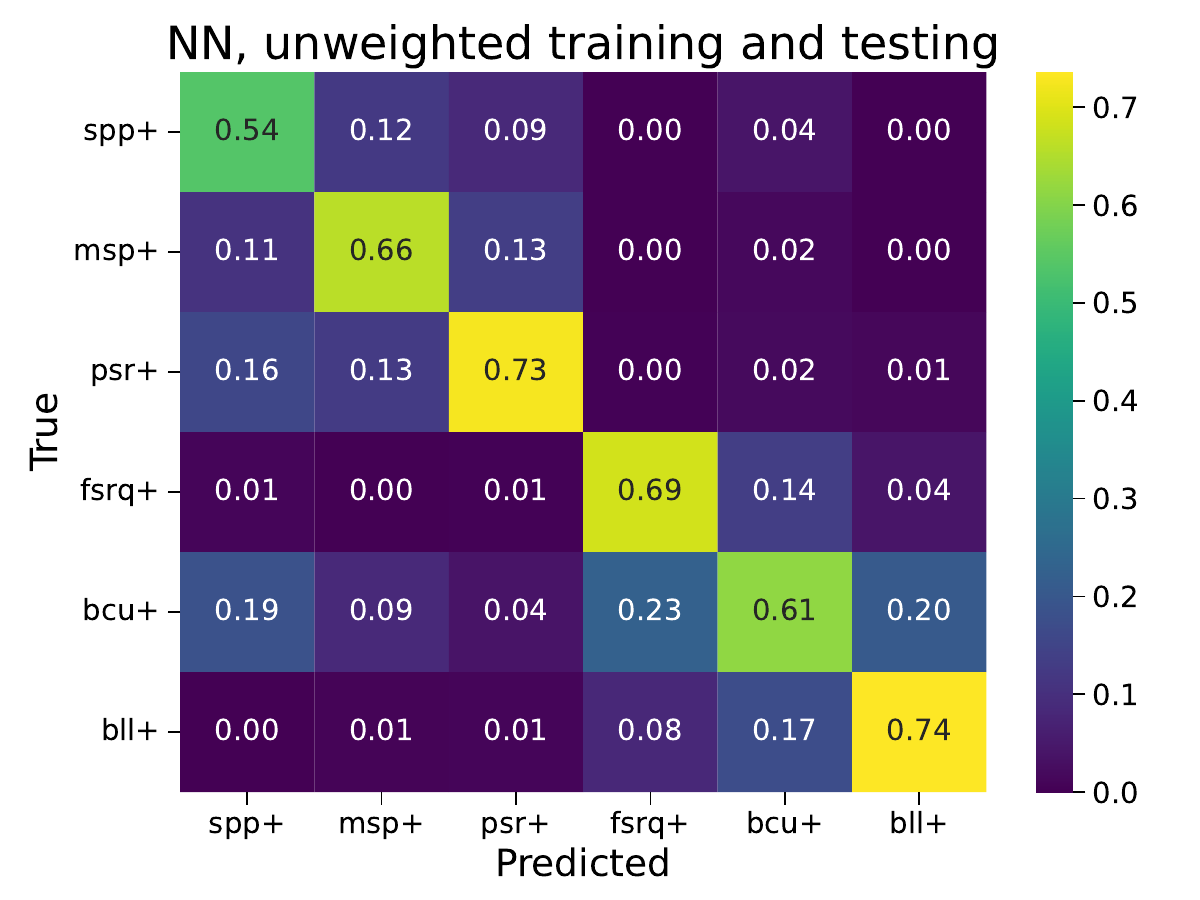}
\includegraphics[width=\cmsize\columnwidth]{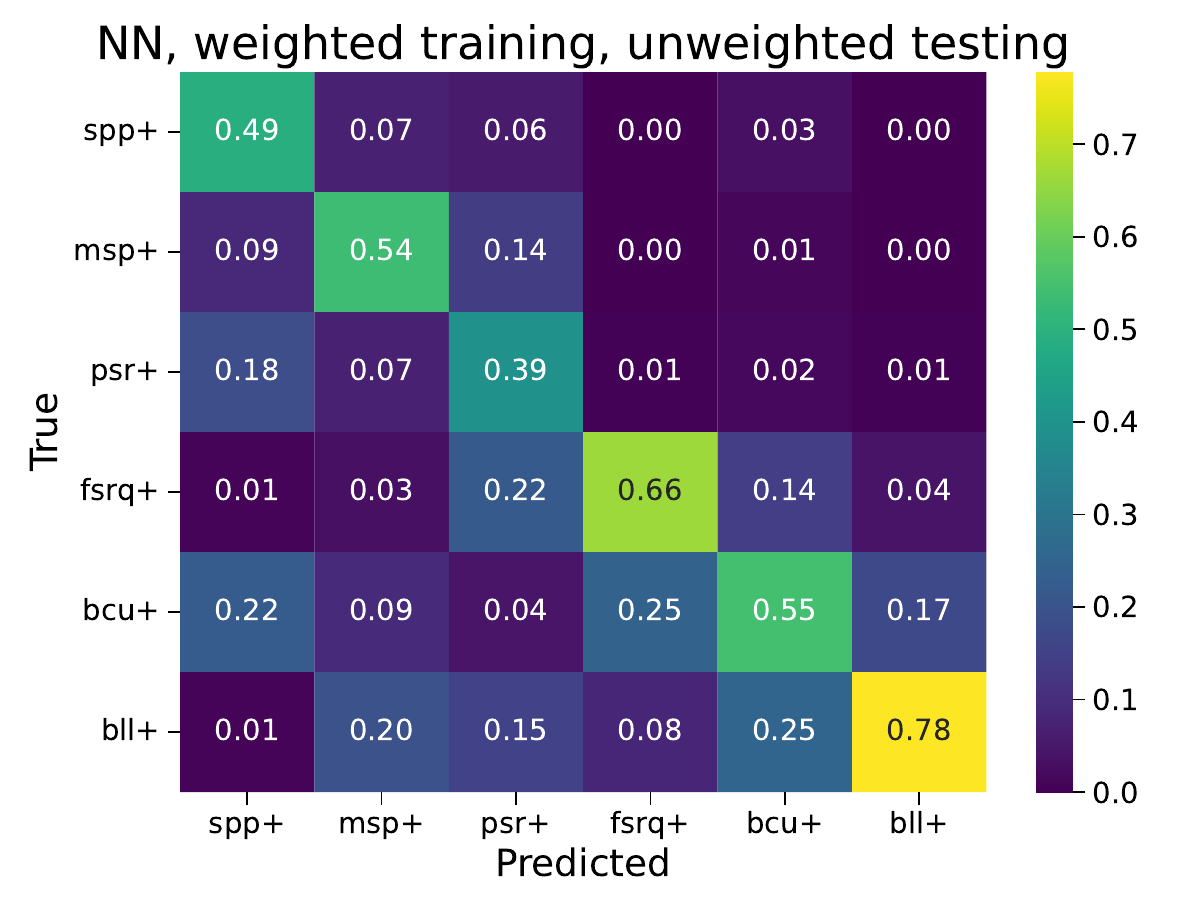} \\
\includegraphics[width=\cmsize\columnwidth]{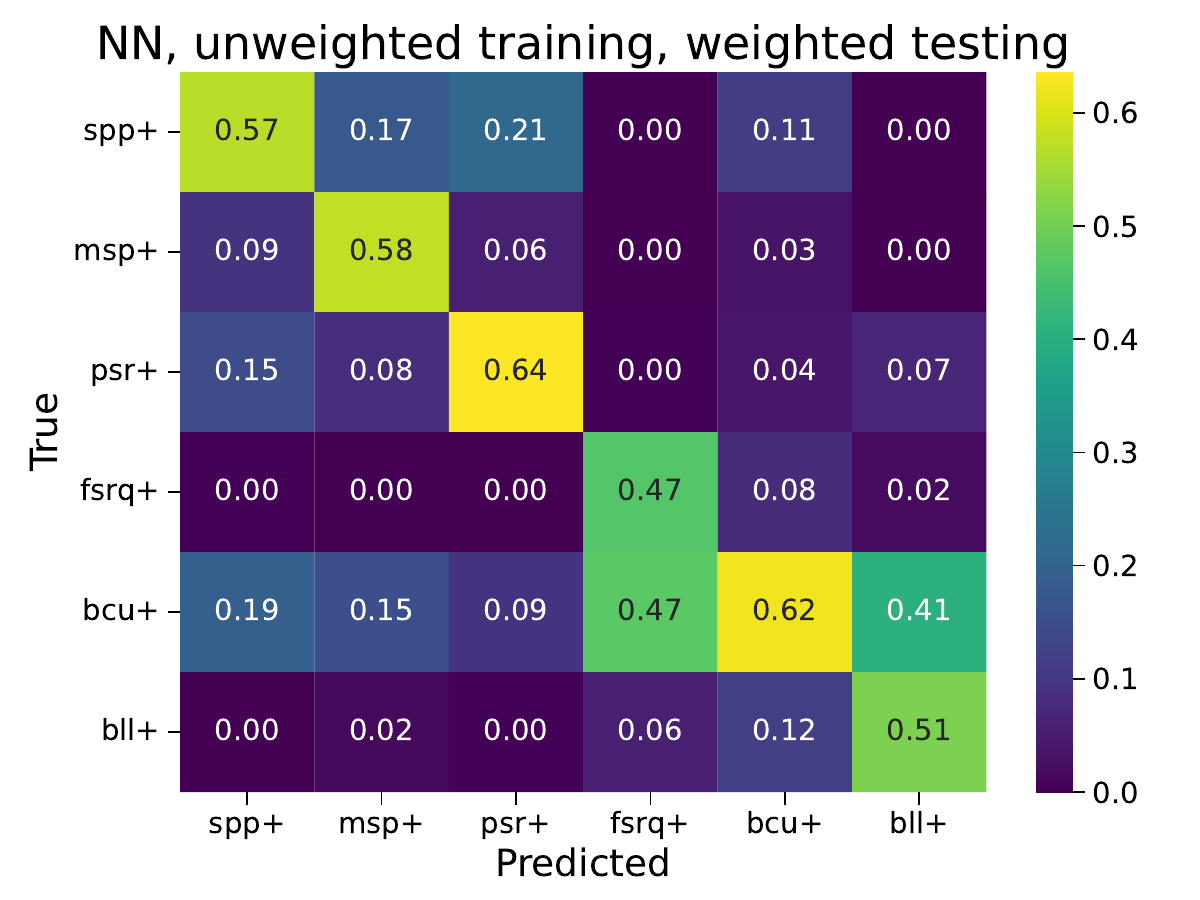}
\includegraphics[width=\cmsize\columnwidth]{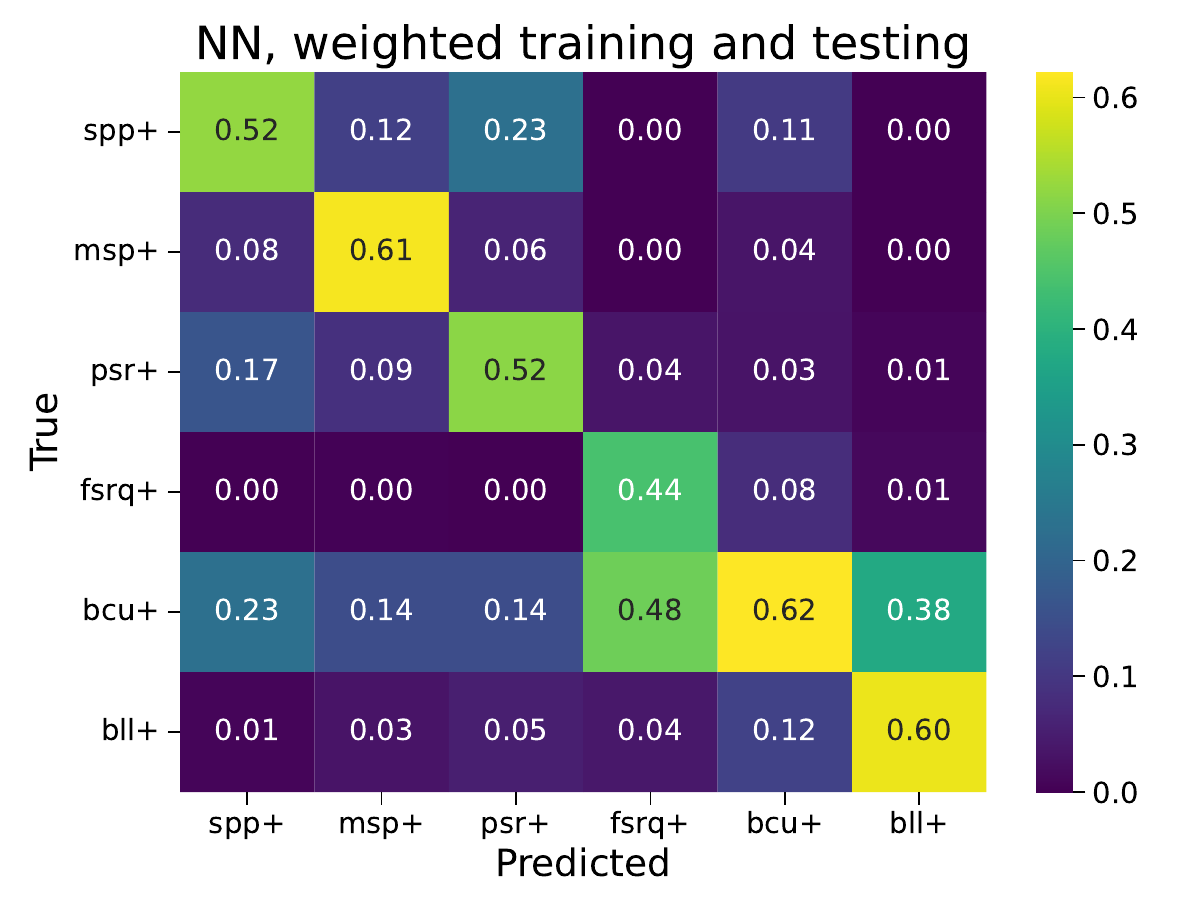}
\caption{
Confusion matrix normalized to the number of predicted sources in a class for the NN classification. The sources are classified according to the highest class probability for each source.
The numbers on the diagonal show the one-vs-all precision for this classification method. Top (bottom) panels show the performance for the unweighted (weighted) samples representative for the associated (unassociated) sources.
}
\label{fig:CM_NN}
\end{figure*}

In Fig. \ref{fig:CM_NN} we compare the confusion matrices for the weighted and unweighted training and testing datasets 
for the classification with the NN algorithm similar to the confusion matrices in Fig. \ref{fig:CM_RF} for the classification with the RF algorithm.
The training on unweighted samples (left panels) has a similar or better performance (with a few exceptions)
than the training on weighted samples (right panels)
when tested both on unweighted samples (top panels) and on weighted samples (bottom panels).

\bsp	
\label{lastpage}
\end{document}